\pdfoutput=1
\documentclass[12pt,a4paper]{article}

\usepackage{comment}
\usepackage{ifthen} 
\newboolean{pdflatex}
\setboolean{pdflatex}{true} 
\newboolean{articletitles}
\setboolean{articletitles}{true} 
\newboolean{uprightparticles}
\setboolean{uprightparticles}{false} 
\def\paperauthors{LHCb collaboration}

\def\Xibsigned        {{\ensuremath{\Xires^{(-,0)}_\bquark}}\xspace}
\def\Xicsigned        {{\ensuremath{\Xires^{(0,+)}_\cquark}}\xspace}
\def\Xibzprime        {{\ensuremath{\Xires'^{0}_\bquark}}\xspace}
\def\Xibmprime        {{\ensuremath{\Xires'^{-}_\bquark}}\xspace}
\def\Xibzstar         {{\ensuremath{\Xires^{*0}_\bquark}}\xspace}
\def\Xibmstar         {{\ensuremath{\Xires^{*-}_\bquark}}\xspace}
\def\Xibprime         {{\ensuremath{\Xires'^{(0,-)}_\bquark}}\xspace}
\def\Xibstar          {{\ensuremath{\Xires^{*(0,-)}_\bquark}}\xspace}
\def\XibzstarstarA    {{\ensuremath{\Xires_\bquark{(6087)}^{0}}}\xspace}
\def\XibmstarstarA    {{\ensuremath{\Xires_\bquark(\mathrm{1P,1/2)^{-}}}}\xspace}
\def\XibzstarstarB    {{\ensuremath{\Xires_\bquark{(6095)}^{0}}}\xspace}
\def\XibmstarstarB    {{\ensuremath{\Xires_\bquark{(6100)}^{-}}}\xspace}
\newboolean{prl}
\setboolean{prl}{False}

\def\paperasciititle{Observation of new baryons in the Xib-pi+pi- and Xib0pi+pi- systems} 
\def\papertitle{Observation of new baryons in the $\Xibm\pip\pim$ and $\Xibz\pip\pim$ systems} 
\def\paperkeywords{{High Energy Physics}, {LHCb}}
\def\papercopyright{\the\year\ CERN for the benefit of the LHCb collaboration} 
\def\paperlicence{CC BY 4.0 licence}
\def\paperlicenceurl{https://creativecommons.org/licenses/by/4.0/}

\usepackage[top=1in, bottom=1.25in, left=1in, right=1in]{geometry}

\usepackage{microtype}
\usepackage{lineno}  
\usepackage{xspace} 
\usepackage{caption} 

\usepackage{graphicx}  
\usepackage{color}
\usepackage{colortbl}
\graphicspath{{./figs/}} 

\usepackage{amsmath} 
\usepackage{amssymb}
\usepackage{amsfonts}
\usepackage{upgreek} 

\newcommand*\patchAmsMathEnvironmentForLineno[1]{%
\expandafter\let\csname old#1\expandafter\endcsname\csname #1\endcsname
\expandafter\let\csname oldend#1\expandafter\endcsname\csname
end#1\endcsname
 \renewenvironment{#1}%
   {\linenomath\csname old#1\endcsname}%
   {\csname oldend#1\endcsname\endlinenomath}%
}
\newcommand*\patchBothAmsMathEnvironmentsForLineno[1]{%
  \patchAmsMathEnvironmentForLineno{#1}%
  \patchAmsMathEnvironmentForLineno{#1*}%
}
\AtBeginDocument{%
\patchBothAmsMathEnvironmentsForLineno{equation}%
\patchBothAmsMathEnvironmentsForLineno{align}%
\patchBothAmsMathEnvironmentsForLineno{flalign}%
\patchBothAmsMathEnvironmentsForLineno{alignat}%
\patchBothAmsMathEnvironmentsForLineno{gather}%
\patchBothAmsMathEnvironmentsForLineno{multline}%
\patchBothAmsMathEnvironmentsForLineno{eqnarray}%
}

\usepackage{hyperxmp}

\usepackage[pdftex,
            pdfauthor={\paperauthors},
            pdftitle={\paperasciititle},
            pdfkeywords={\paperkeywords},
            pdfcopyright={Copyright (C) \papercopyright},
            pdflicenseurl={\paperlicenceurl}]{hyperref}

\usepackage[bottom,flushmargin,hang,multiple]{footmisc}

\usepackage[all]{hypcap} 

\usepackage{xspace} 
\usepackage{upgreek}

\def\lhcb   {\mbox{LHCb}\xspace}

\def\cms    {\mbox{CMS}\xspace}

\def\MagUp {\mbox{\em Mag\kern -0.05em Up}\xspace}

\ifthenelse{\boolean{uprightparticles}}%
{

 \def\Ppi         {\ensuremath{\uppi}\xspace}

 \def\Ppsi        {\ensuremath{\uppsi}\xspace}

 \def\PDelta      {\ensuremath{\Delta}\xspace}                 
 \def\PXi         {\ensuremath{\Xi}\xspace}                 
 \def\PLambda     {\ensuremath{\Lambda}\xspace}                 
 \def\PSigma      {\ensuremath{\Sigma}\xspace}                 
 \def\POmega      {\ensuremath{\Omega}\xspace}                 
 \def\PUpsilon    {\ensuremath{\Upsilon}\xspace}
 \let\oldPi\Pi
 \def\PPi         {\ensuremath{\oldPi}\xspace}

 \def\PB      {\ensuremath{\mathrm{B}}\xspace}                 
                  
 \def\PD      {\ensuremath{\mathrm{D}}\xspace}

 \def\PJ      {\ensuremath{\mathrm{J}}\xspace}                 
 \def\PK      {\ensuremath{\mathrm{K}}\xspace}

 \def\Pb      {\ensuremath{\mathrm{b}}\xspace}                 
 \def\Pc      {\ensuremath{\mathrm{c}}\xspace}

 \def\Pi      {\ensuremath{\mathrm{i}}\xspace}

 \def\Ps      {\ensuremath{\mathrm{s}}\xspace}

 \def\thebaroffset{0.0em}
}
{

 \def\Ppi         {\ensuremath{\pi}\xspace}

 \def\Ppsi        {\ensuremath{\psi}\xspace}                 
                  
 \mathchardef\PDelta="7101
 \mathchardef\PXi="7104
 \mathchardef\PLambda="7103
 \mathchardef\PSigma="7106
 \mathchardef\POmega="710A
 \mathchardef\PUpsilon="7107
 \mathchardef\PPi="7105
                  
 \def\PB      {\ensuremath{B}\xspace}                 
                  
 \def\PD      {\ensuremath{D}\xspace}

 \def\PJ      {\ensuremath{J}\xspace}                 
 \def\PK      {\ensuremath{K}\xspace}

 \def\Pb      {\ensuremath{b}\xspace}                 
 \def\Pc      {\ensuremath{c}\xspace}

 \def\Pi      {\ensuremath{i}\xspace}

 \def\Ps      {\ensuremath{s}\xspace}

 \def\thebaroffset{0.18em}
}
\newcommand{\offsetoverline}[2][\thebaroffset]{\kern #1\overline{\kern -#1 #2}}%

\makeatletter
\ifcase \@ptsize \relax
  \newcommand{\miniscule}{\@setfontsize\miniscule{4}{5}}
\or
  \newcommand{\miniscule}{\@setfontsize\miniscule{5}{6}}
\or
  \newcommand{\miniscule}{\@setfontsize\miniscule{5}{6}}
\fi
\makeatother

\DeclareRobustCommand{\optbar}[1]{\shortstack{{\miniscule (\rule[.5ex]{1.25em}{.18mm})}
  \\ [-.7ex] $#1$}}

\def\squark    {{\ensuremath{\Ps}}\xspace}

\def\cquark    {{\ensuremath{\Pc}}\xspace}

\def\bquark    {{\ensuremath{\Pb}}\xspace}

\def\pion   {{\ensuremath{\Ppi}}\xspace}
\def\piz    {{\ensuremath{\pion^0}}\xspace}
\def\pip    {{\ensuremath{\pion^+}}\xspace}
\def\pim    {{\ensuremath{\pion^-}}\xspace}

\def\kaon    {{\ensuremath{\PK}}\xspace}

\def\KorKbar {\kern \thebaroffset\optbar{\kern -\thebaroffset \PK}{}\xspace}

\def\Km      {{\ensuremath{\kaon^-}}\xspace}

\def\D       {{\ensuremath{\PD}}\xspace}

\def\DorDbar {\kern \thebaroffset\optbar{\kern -\thebaroffset \PD}\xspace}
\def\Dz      {{\ensuremath{\D^0}}\xspace}

\def\Dp      {{\ensuremath{\D^+}}\xspace}
\def\Dm      {{\ensuremath{\D^-}}\xspace}

\def\DpDm    {\ensuremath{\Dp {\kern -0.16em \Dm}}\xspace}

\def\Ds      {{\ensuremath{\D^+_\squark}}\xspace}

\def\B       {{\ensuremath{\PB}}\xspace}

\def\BorBbar {\kern \thebaroffset\optbar{\kern -\thebaroffset \PB}\xspace}

\def\Bd      {{\ensuremath{\B^0}}\xspace}

\def\BdorBdbar {\kern \thebaroffset\optbar{\kern -\thebaroffset \Bd}\xspace}

\def\Bs      {{\ensuremath{\B^0_\squark}}\xspace}

\def\BsorBsbar {\kern \thebaroffset\optbar{\kern -\thebaroffset \Bs}\xspace}

\def\jpsi     {{\ensuremath{{\PJ\mskip -3mu/\mskip -2mu\Ppsi}}}\xspace}

\def\Y#1S{\ensuremath{\PUpsilon{(#1S)}}\xspace}

\def\Lz          {{\ensuremath{\PLambda}}\xspace}

\def\LorLbar     {\kern \thebaroffset\optbar{\kern -\thebaroffset \PLambda}\xspace}

\def\Xires       {{\ensuremath{\PXi}}\xspace}

\def\Lc          {{\ensuremath{\Lz^+_\cquark}}\xspace}

\def\Xic         {{\ensuremath{\Xires_\cquark}}\xspace}
\def\Xicz        {{\ensuremath{\Xires^0_\cquark}}\xspace}
\def\Xicp        {{\ensuremath{\Xires^+_\cquark}}\xspace}

\def\Xib          {{\ensuremath{\Xires_\bquark}}\xspace}
\def\Xibz         {{\ensuremath{\Xires^0_\bquark}}\xspace}
\def\Xibm         {{\ensuremath{\Xires^-_\bquark}}\xspace}

\def\to                 {\ensuremath{\rightarrow}\xspace}

\def\AT#1     {\ensuremath{A_{\mathrm{T}}^{#1}}\xspace}           

\def\C#1      {\ensuremath{\mathcal{C}_{#1}}\xspace}                       
\def\Cp#1     {\ensuremath{\mathcal{C}_{#1}^{'}}\xspace}                    
\def\Ceff#1   {\ensuremath{\mathcal{C}_{#1}^{\mathrm{(eff)}}}\xspace}        
\def\Cpeff#1  {\ensuremath{\mathcal{C}_{#1}^{'\mathrm{(eff)}}}\xspace}       
\def\Ope#1    {\ensuremath{\mathcal{O}_{#1}}\xspace}                       
\def\Opep#1   {\ensuremath{\mathcal{O}_{#1}^{'}}\xspace}                    

\newcommand{\aunit}[1]{\ensuremath{\text{\,#1}}}       

\newcommand{\tev}{\aunit{Te\kern -0.1em V}\xspace}
\newcommand{\gev}{\aunit{Ge\kern -0.1em V}\xspace}
\newcommand{\mev}{\aunit{Me\kern -0.1em V}\xspace}
\newcommand{\kev}{\aunit{ke\kern -0.1em V}\xspace}
\newcommand{\ev}{\aunit{e\kern -0.1em V}\xspace}
 
\newcommand{\mevc}{\ensuremath{\aunit{Me\kern -0.1em V\!/}c}\xspace}
\newcommand{\gevc}{\ensuremath{\aunit{Ge\kern -0.1em V\!/}c}\xspace}
\newcommand{\mevcc}{\ensuremath{\aunit{Me\kern -0.1em V\!/}c^2}\xspace}
\newcommand{\gevcc}{\ensuremath{\aunit{Ge\kern -0.1em V\!/}c^2}\xspace}

\def\fb   {\ensuremath{\aunit{fb}}\xspace}
\def\invfb   {\ensuremath{\fb^{-1}}\xspace}

\def\gsim{{~\raise.15em\hbox{$>$}\kern-.85em
          \lower.35em\hbox{$\sim$}~}\xspace}
\def\lsim{{~\raise.15em\hbox{$<$}\kern-.85em
          \lower.35em\hbox{$\sim$}~}\xspace}

\def\tell1  {TELL1\xspace}
\def\ukl1   {UKL1\xspace}

\newcommand{\eg}{\mbox{\itshape e.g.}\xspace}

\newcommand{\lhcborcid}[1]{\href{https://orcid.org/#1}{\hspace*{0.1em}\raisebox{-0.45ex}{\includegraphics[width=1em]{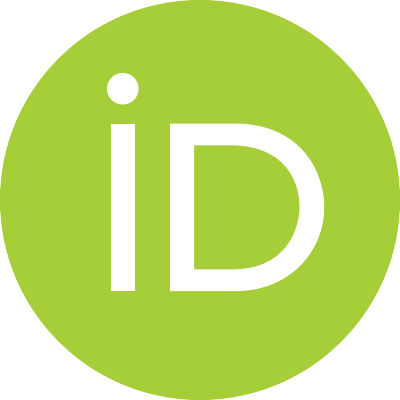}}}}

\usepackage{cite} 
\usepackage{mciteplus}
\begin{document}
\renewcommand{\thefootnote}{\fnsymbol{footnote}}
\setcounter{footnote}{1}

\begin{titlepage}
\pagenumbering{roman}

\vspace*{-1.5cm}
\centerline{\large EUROPEAN ORGANIZATION FOR NUCLEAR RESEARCH (CERN)}
\vspace*{1.5cm}
\noindent
\begin{tabular*}{\linewidth}{lc@{\extracolsep{\fill}}r@{\extracolsep{0pt}}}
\ifthenelse{\boolean{pdflatex}}
{\vspace*{-1.5cm}\mbox{\!\!\!\includegraphics[width=.14\textwidth]{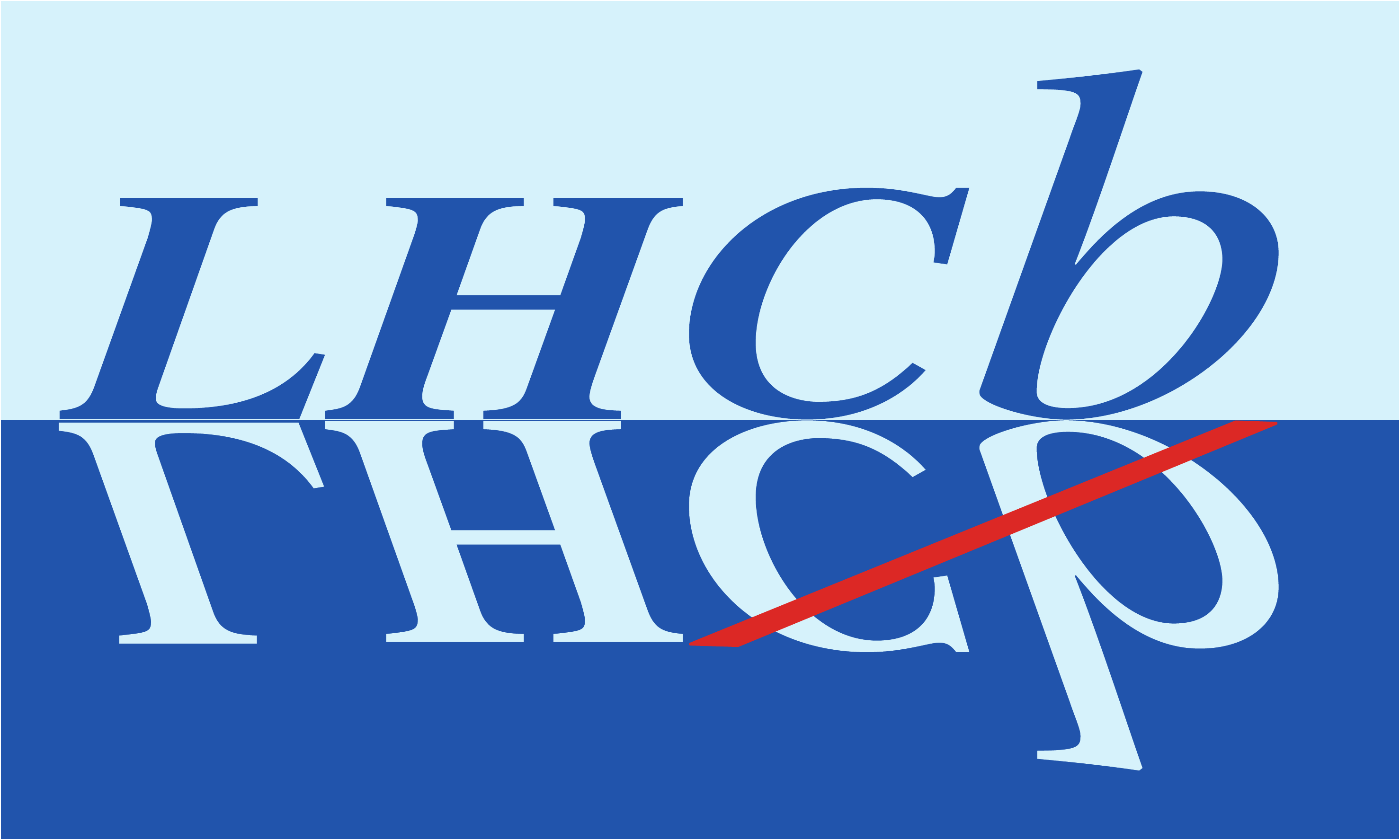}} & &}%
{\vspace*{-1.2cm}\mbox{\!\!\!\includegraphics[width=.12\textwidth]{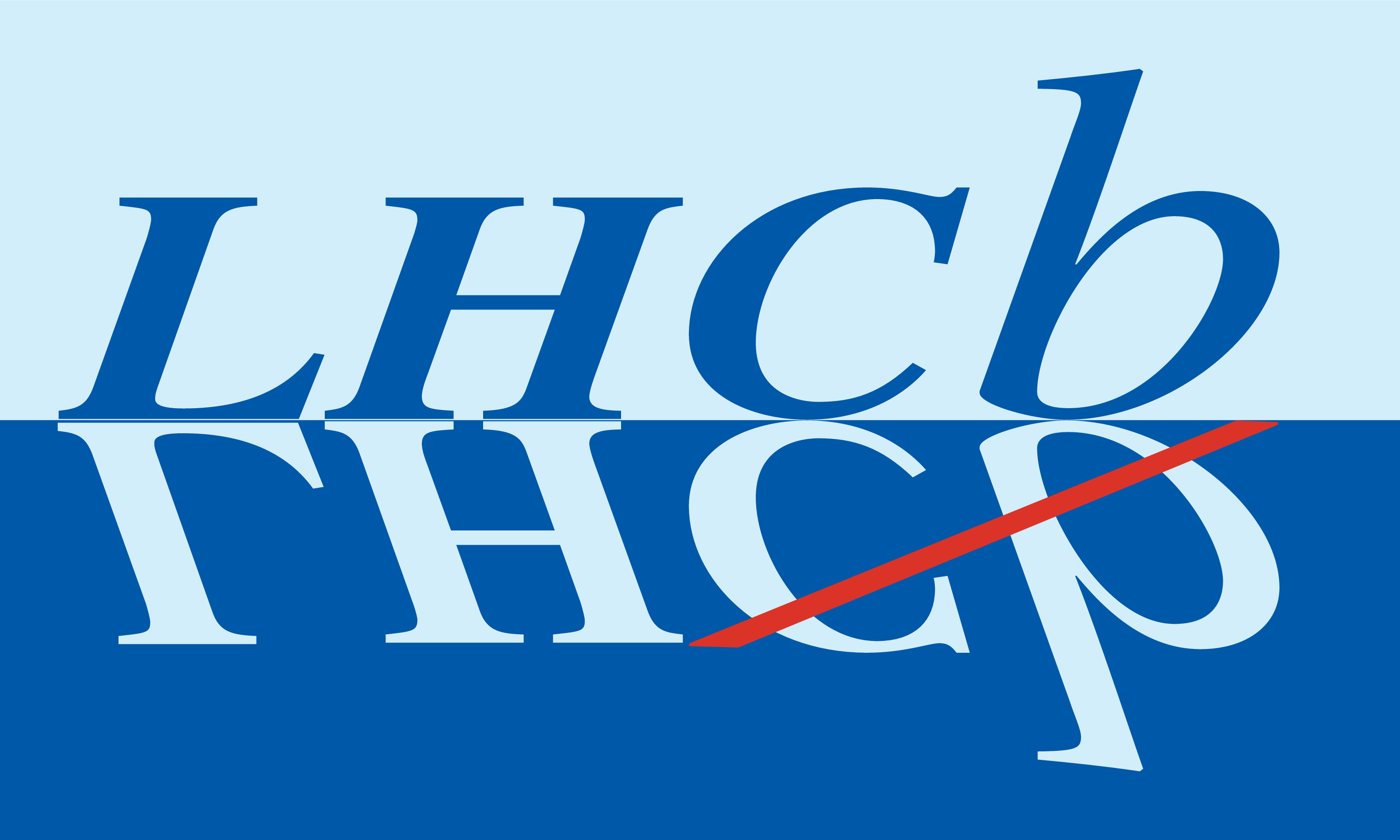}} & &}%
\\
 & & CERN-EP-2023-126 \\  
 & & LHCb-PAPER-2023-008 \\  
 & & \today \\ 
 & & \\
\end{tabular*}

\vspace*{4.0cm}

{\normalfont\bfseries\boldmath\huge
\begin{center}
  \papertitle 
\end{center}
}

\vspace*{2.0cm}

\begin{center}
\paperauthors\footnote{Authors are listed at the end of this Letter.}
\end{center}

\vspace{\fill}

\begin{abstract}
\noindent
The first observation and study of two new baryonic structures in the final state $\Xibz\pip\pim$ and the confirmation of the \XibmstarstarB state in the $\Xibm\pip\pim$ decay mode are reported using proton-proton
collision data collected by the \lhcb experiment, corresponding to an integrated luminosity of 9\invfb. In addition, the properties of the 
known \Xibzstar, \Xibmprime and \Xibmstar resonances are measured with  improved precision. The new decay mode of the \Xibz baryon to the \Xicp\pim\pip\pim final state is observed and exploited for the first time in these measurements.

\end{abstract}

\vspace*{2.0cm}

\begin{center}
Published in 
\href{https://journals.aps.org/prl/abstract/10.1103/PhysRevLett.131.171901}{Phys. Rev. Lett. 131 (2023) 171901}
\end{center}

\vspace{\fill}

{\footnotesize 
\centerline{\copyright~\papercopyright. \href{\paperlicenceurl}{\paperlicence}.}}
\vspace*{2mm}

\end{titlepage}


\newpage
\setcounter{page}{2}
\mbox{~}

\renewcommand{\thefootnote}{\arabic{footnote}}
\setcounter{footnote}{0}
\cleardoublepage

\pagestyle{plain} 
\setcounter{page}{1}
\pagenumbering{arabic}
The \Xibsigned baryons form an isospin doublet and are made of a $b$ quark,
an $s$ quark and a lighter $q$ ($u$ or $d$) quark.
Their ground states have angular momentum $L=0$ between the $b$ quark and the light diquark.
Three isospin doublets of such non-excited states are expected in the
quark model~\cite{Gell-Mann:1964ewy,Zweig:352337,Klempt:2009pi}
with different spin-parity $J^P$ and angular momentum of the $sq$ diquark $J_{sq}$.
The \Xibsigned, \Xibprime and \Xibstar states are characterized by  ($J_{sq}$, $J^P$) values of ($0,{\frac{1}{2}}^+$), ($1,{\frac{1}{2}}^+$) and ($1, {\frac{3}{2}}^+$), respectively.
Although five of these states have been observed
experimentally~\cite{PhysRevLett.99.052002,
CDF:2011ipk,
CMS:2012frl,
LHCb:2014nae},
the \Xibzprime baryon remains unobserved.
This may be because its mass lies below the threshold for the
$\Xibzprime\to\Xibm\pip$ decay~\cite{LHCb-PAPER-2020-016,PhysRevD.96.116016},
meaning that it only decays to either the $\Xibz\piz$ or $\Xibz\gamma$ final states making it experimentally challenging  to observe.
A number of excited states of higher mass is expected, with predictions for their properties available, \eg in Refs.~\cite{PhysRevD.98.031502,PhysRevD.99.094016,PhysRevD.96.116016,PhysRevD.104.054012,th2_2021,PhysRevD.103.094003,Chen:2014nyo,PhysRevD.84.014025,Roberts:2007ni}.
The \cms collaboration has reported the observation of the \XibmstarstarB resonance in the $\Xibm\pip\pim$ final state,
using \Xibm decays to final states containing \jpsi mesons~\cite{PhysRevLett.126.252003}.

In this Letter,
both the \Xibm\pip\pim and \Xibz\pip\pim final states and their intermediate \Xibm\pip and \Xibz\pim states
are investigated experimentally
(the inclusion of charge conjugate processes and the use of natural units are implicit throughout this Letter),
using $pp$ collision data collected by the \lhcb experiment
at centre-of-mass energies of $7,8,13$ \tev, corresponding to an integrated luminosity of 9\invfb.
The observation of three narrow states is reported and their properties are measured.

The \lhcb detector~\cite{LHCb-DP-2008-001,LHCb-DP-2014-002} is a single-arm forward spectrometer
covering the \mbox{pseudorapidity} range $2<\eta <5$.
The detector includes a high-precision tracking system consisting of a silicon-strip vertex detector
surrounding the $pp$ interaction region~\cite{LHCb-DP-2014-001},
a large-area silicon-strip detector located upstream of a dipole magnet
with a bending power of about $4{\mathrm{\,Tm}}$, and three stations of silicon-strip detectors and straw
drift tubes~\cite{LHCb-DP-2013-003,LHCb-DP-2017-001} placed downstream of the magnet.
Different types of charged hadrons are distinguished using information
from two ring-imaging Cherenkov detectors~\cite{LHCb-DP-2012-003}.
Simulated data samples are produced with the software packages described in Refs.~\cite{Sjostrand:2007gs,LHCb:2011dpk,Lange:2001uf,Allison:2006ve,Clemencic_2011} and are used to model
the detector resolution and optimize the selection criteria.

Samples of \Xibm (\Xibz) candidates are formed from $\Xicz\pim$ ($\Xicp\pim$) and
$\Xicz\pim\pip\pim$ ($\Xicp\pim\pip\pim$) combinations,
where the \Xicz (\Xicp) baryon is reconstructed in the $p \Km \Km \pip$ ($p \Km \pip$) final state.
To suppress background coming from promptly produced particles, all \Xibsigned decay products
are required to be displaced significantly from 
all primary $pp$ collision vertices (PVs) in the event.
The reconstructed \Xicz (\Xicp) candidates are required to have a mass within 20\mev (25\mev) of the
respective world-average mass values~\cite{PDG2022}.
Displaced pion tracks are combined with \Xic candidates to form \Xibsigned candidates, 
requiring good vertex-fit quality and significant displacement of the \Xibsigned decay point from any PV in the event.
All charged particles are required to have particle-identification (PID) information consistent with their respective mass hypotheses.
PID variables are based on neural network algorithms~\cite{DeCian:2255039} and their distributions
in simulation are calibrated 
using data~\cite{LHCb-DP-2018-001}.
If multiple candidates per collision event pass the selection requirements, all of them are preserved in the data sample.
The topological, kinematic and PID variables are used as inputs to a 
Boosted Decision Tree (BDT) classifier~\cite{ROE2005577} that discriminates  \Xibsigned signal candidates from random track combinations.
The classifier is trained using simulated \Xibsigned decays as a signal proxy 
and \Xibsigned data candidates in the sideband $5900 < m(\Xicsigned\pim,\Xicsigned\pim\pip\pim) < 6000\mev$ as a background proxy.
The mass distributions of the selected $\Xibsigned \rightarrow \Xicsigned \pim$ and $\Xibsigned \rightarrow \Xicsigned \pim\pip\pim$
candidates are shown in the supplemental material~\cite{supplemental}.
The decay mode $\Xibz \rightarrow \Xicp\pim\pip\pim$ is observed for the first time experimentally.
The selection requirement on the BDT classifier output is optimized for the observation of new states
and retains 96\% of the \Xibm and  92\% of the \Xibz signal candidates.
Additional vetoes  are imposed, as described in Ref.~\cite{LHCb-PAPER-2019-042}, to suppress other abundant processes with displaced vertices \eg those coming from \Dz, \Dp, \Ds and \Lc decays with a misidentified particle.

The \Xibsigned candidates within a $\pm 75$\mev window around the peak position are combined
with one charged pion (two pions) to investigate the
\Xibm\pip and \Xibz\pim (\Xibm\pip\pim and \Xibz\pim\pip)
mass spectra.
In order to improve the mass resolution, the
obtained candidates are refitted
with the masses of the \Xibsigned and \Xicsigned baryon candidates constrained to their known values~\cite{PDG2022}
and the \Xibsigned flight direction
to originate from the PV~\cite{Hulsbergen:2005pu}.
Additional requirements are applied to the $\Xib\pim\pip$ candidates, where 
the \Xibzstar, \Xibmprime and \Xibmstar intermediate states are selected
according to their observed widths and known mass values~\cite{PDG2022}.
Signal mass windows for the $\Xib\pi$ intermediate resonances are defined as $|m(\Xib \pi)-m_{\Xibzstar}|<3\mev$,
$|m(\Xib \pi)-m_{\Xibmprime}|<1.25\mev$
and $|m(\Xib \pi)-m_{\Xibmstar}|<5\mev$,
each corresponding to 2.5\,$\sigma$ of the observed experimental peak.

\begin{figure}[t]
    \centering
    \ifthenelse{\boolean{prl}}
    {\includegraphics[width=1.0\linewidth]{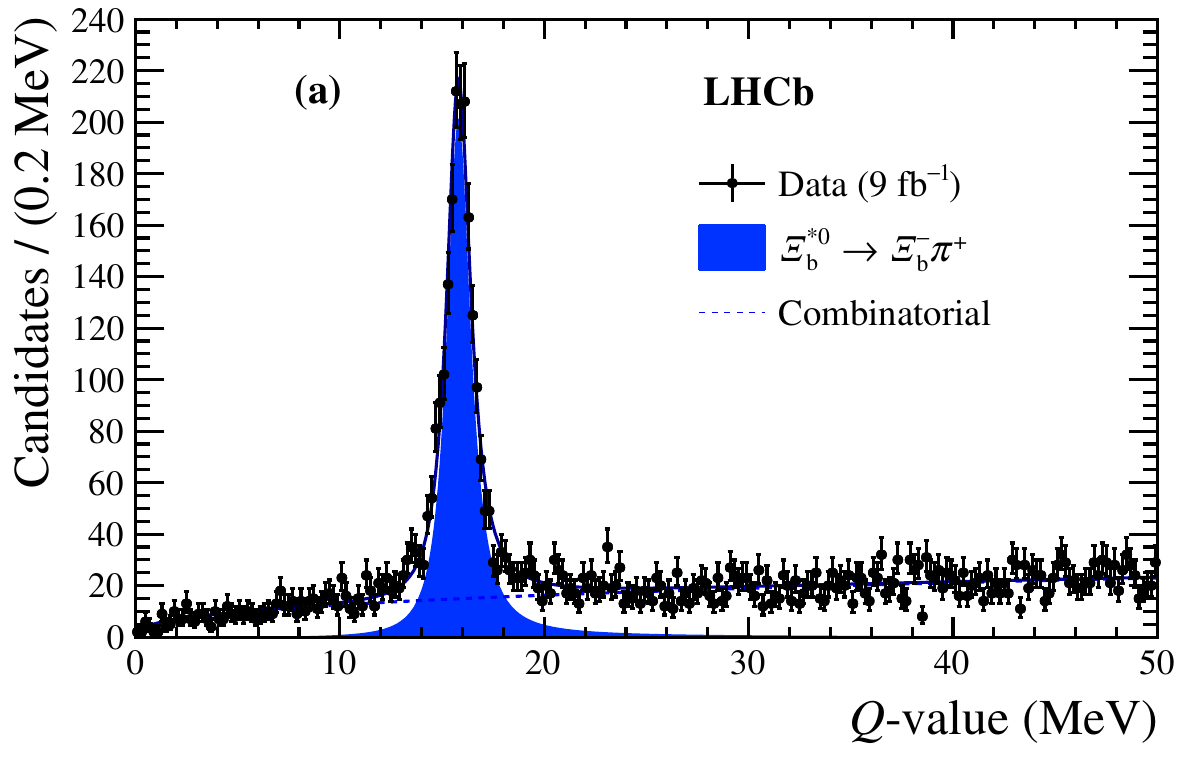}}
    {\includegraphics[width=0.6\linewidth]{fig1a.pdf}}
    \ifthenelse{\boolean{prl}}
    {\includegraphics[width=1.0\linewidth]{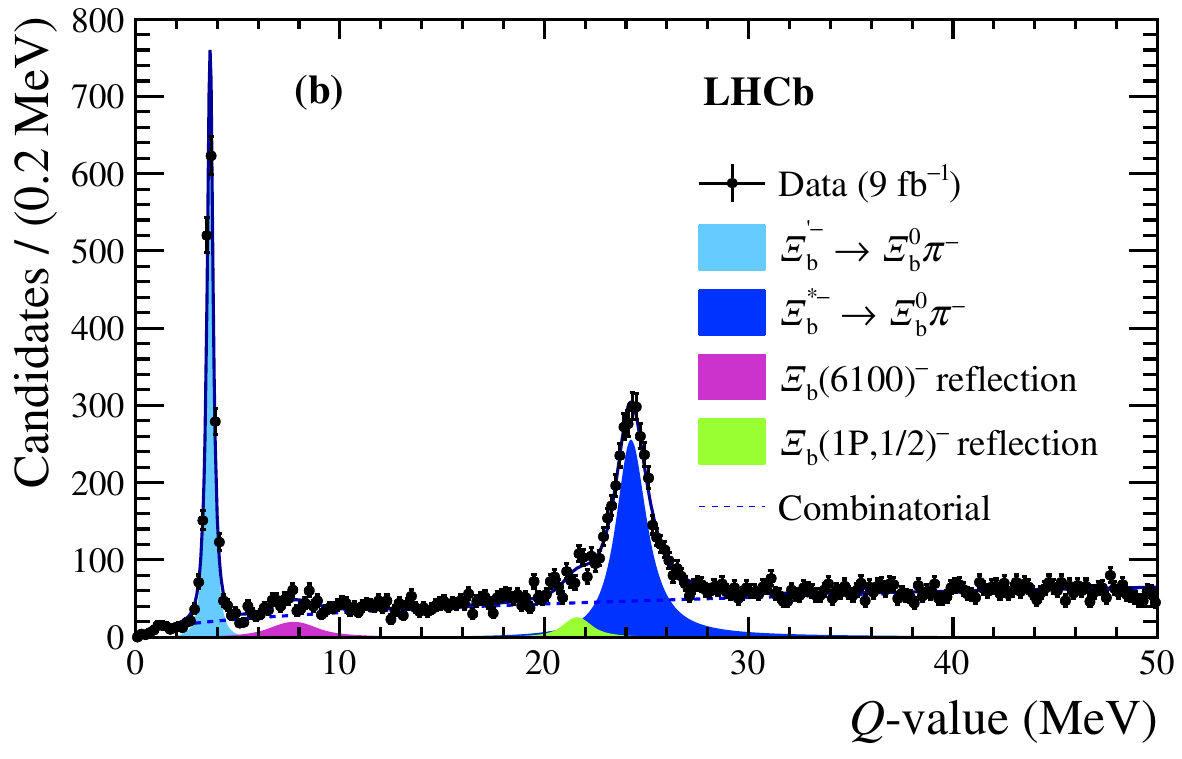}}
    {\includegraphics[width=0.6\linewidth]{fig1b.pdf}}
    \caption{\small Distributions of the mass difference $Q = m_{\Xib\pi} - m_{\Xib} - m_{\pi}$ for selected (a) \Xibm\pip and (b) \Xibz\pim  candidates. The fit results are superimposed.\label{fig:1}}
\end{figure}

\begin{figure}[t]
    \centering
    \ifthenelse{\boolean{prl}}
    {\includegraphics[width=1.0\linewidth]{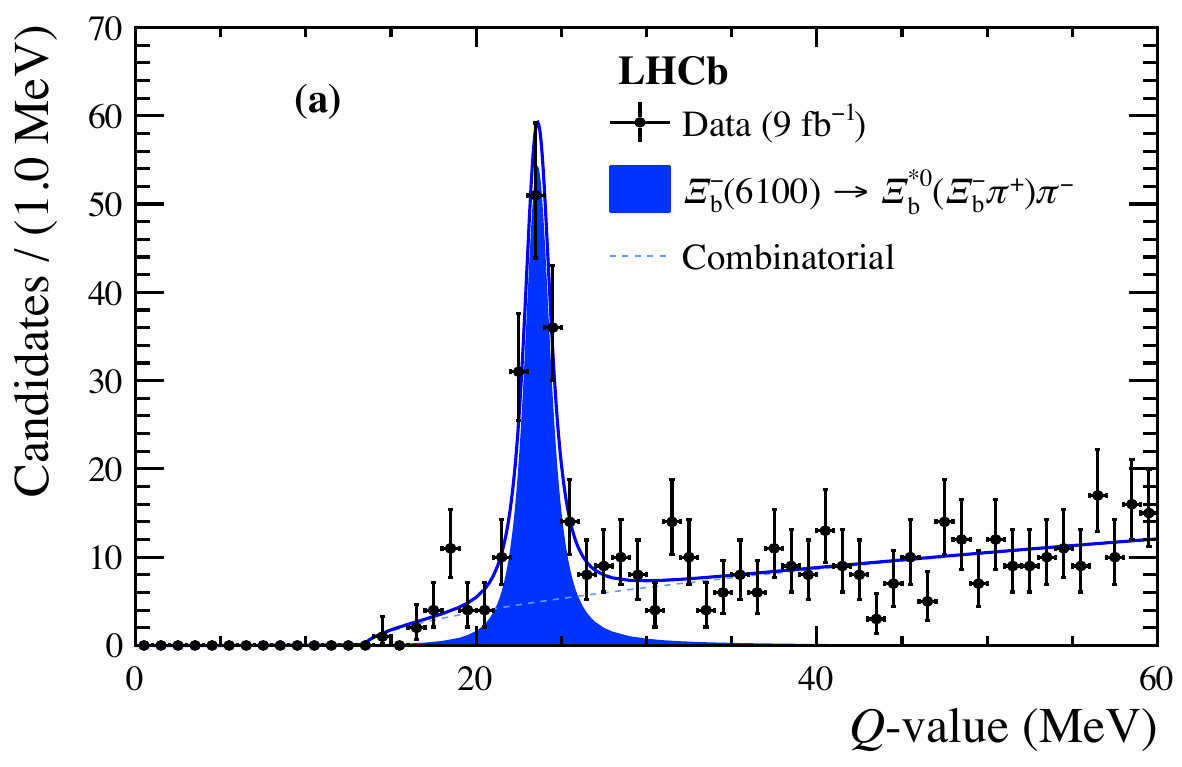}}
    {\includegraphics[width=0.6\linewidth]{fig2a.pdf}}
    \ifthenelse{\boolean{prl}}
    {\includegraphics[width=1.0\linewidth]{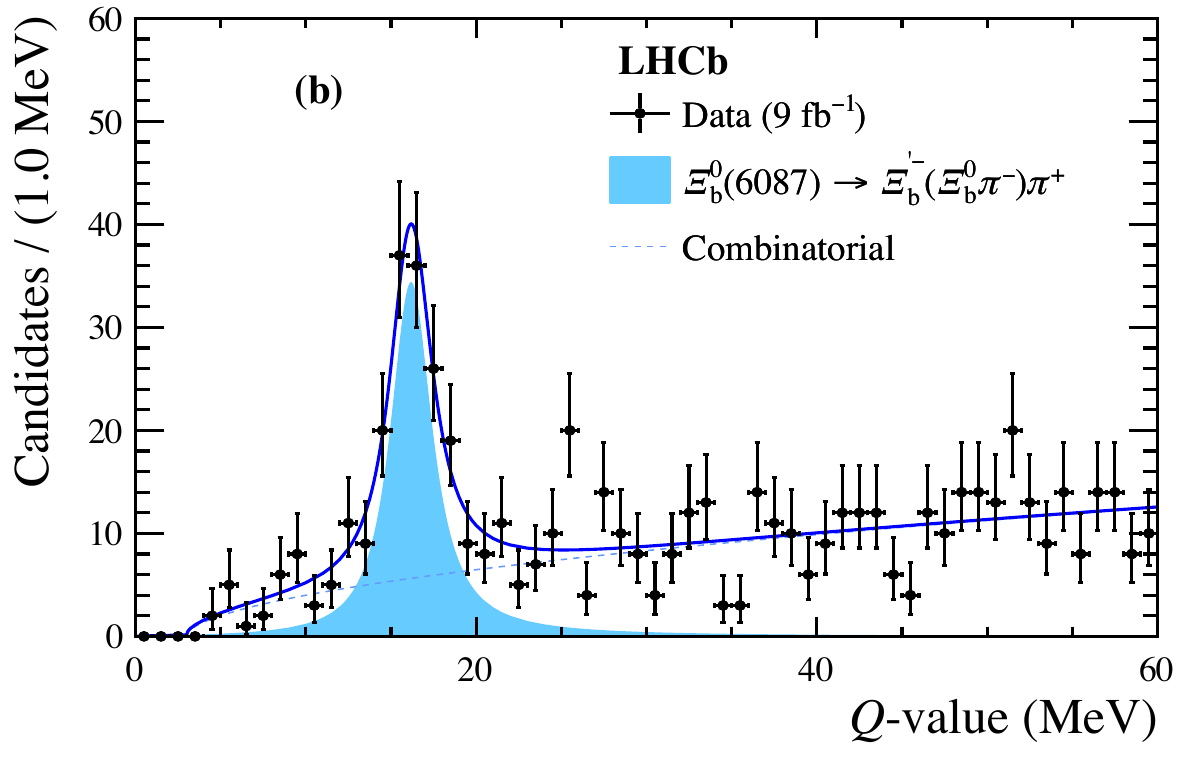}}
    {\includegraphics[width=0.6\linewidth]{fig2b.pdf}}
    \ifthenelse{\boolean{prl}}
    {\includegraphics[width=1.0\linewidth]{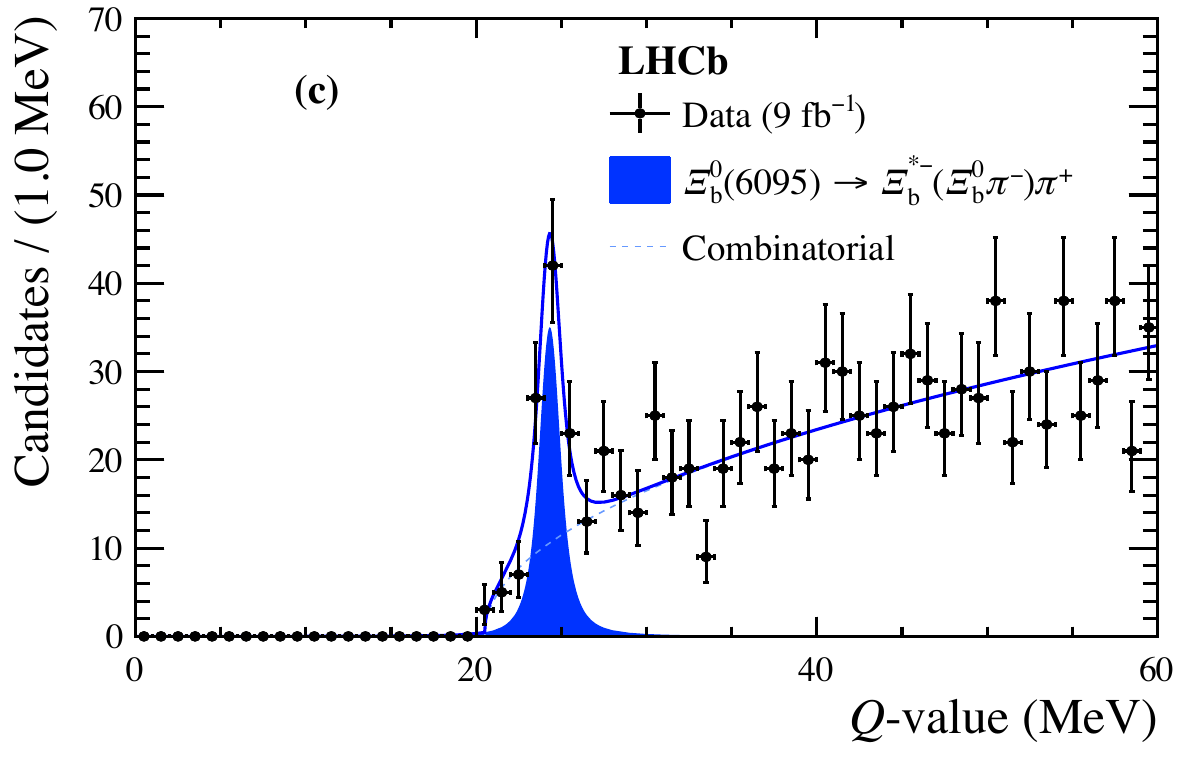}}
    {\includegraphics[width=0.6\linewidth]{fig2c.pdf}}
    \caption{\small Distributions of the mass difference $Q = m_{\Xib\pi\pi} - m_{\Xib} - 2 m_{\pi}$ for selected
    (a) \Xibm\pip\pim candidates in the \Xibzstar mass window, 
    (b) \Xibz\pim\pip candidates in the \Xibmprime mass window and
    (c) \Xibz\pim\pip candidates in the \Xibmstar mass window.
    The fit results are superimposed.\label{fig:2}}
\end{figure}

The signal yields and lineshape parameters of the signal resonances are determined with an extended unbinned maximum-likelihood
fit to the $Q$-value distributions defined as
$m_{\Xib\pi} - m_\Xib - m_\pi$ and $m_{\Xib\pi\pi} - m_\Xib - 2m_\pi$
for $\Xib\pi$ and $\Xib\pi\pi$ decays, respectively.
The mass distributions of the
$\Xibm\pip$ and $\Xibz\pim$ ($\Xibm\pip\pim$ and $\Xibz\pim\pip$) samples are shown in Fig.~\ref{fig:1}~(Fig.~\ref{fig:2}), together
with the results of the fit.
All signal components are modeled using relativistic Breit--Wigner distributions~\cite{Jackson1964}
including Blatt--Weisskopf form factors~\cite{Blatt:1952ije} with a radius of $3\gev^{-1}$.
The orbital angular momentum between the \Xibsigned baryons and the pions is assumed according to the expected spin assignment.
The relativistic Breit--Wigner distributions are convolved with functions parameterizing
the detector resolution.
These resolution models are determined from simulation samples and are consistent with a resolution that scales as $\sqrt{Q}$.
Simulation shows that for each resonance, the resolution is comparable to or smaller than the measured natural widths of the peaks,
with the exception of the \Xibmprime baryon.
The background contribution is parameterized as $(Q - d)^n$, where $d$ and $n$
parameters vary freely in the fit.
This function, which is validated using wrong-charge $\Xibm\pim$ and $\Xibz\pip$ candidates, is sufficient to describe the smooth background coming from random track combinations.
Additional components are included in the fit to the \Xibz\pim spectrum  to describe 
partially reconstructed candidates coming from higher mass resonances.
These components are referred to as \emph{reflections} in the rest of this Letter.
The reflections of the newly observed states in the $\Xib\pi\pi$ mode to the $\Xib\pi$ spectrum
are studied with simulation
and dedicated components are included in the fit, modeled as Gaussian functions with power-law tails~\cite{Skwarnicki:1986xj}.
The means of the reflection components vary freely in the fits to data and their fitted values are consistent
with expectations from simulated backgrounds and cross-checks in data.
The fit confirms the presence of partially reconstructed
$\XibmstarstarB \rightarrow \Xibzstar(\Xibz\pi^0)\pim$ decays and shows hints of a contribution from the decay chain $\XibmstarstarA \rightarrow \Xibzprime(\Xibz\pi^0)\pim$, where 
neither the \XibmstarstarA state, the expected lighter resonance equivalent to that found in the neutral system, nor the \Xibzprime state has been observed experimentally to date (Fig.~\ref{fig:2}).
This component has been validated simulating different mass and width hypotheses for the two particles involved,
taking into account expected isospin splittings given the masses of their charged partners.
However, a precise estimation of the \XibmstarstarA and \Xibzprime state properties is not 
possible due to the limited yield and the presence of two unknown mass values.
The resolutions of the signal components are fixed to the values obtained from simulation.
The fit models are validated with \mbox{pseudoexperiments} and no significant bias is found on any of the parameters of interest.

The fitted yields in the $\Xib\pi$ mass spectra
are $2019 \pm 58$ for the \Xibzstar baryon,
$1750 \pm 50$ for the \Xibmprime baryon and $3380 \pm 110$ for the \Xibmstar baryon.
The \XibmstarstarB state~\cite{PhysRevLett.126.252003} is confirmed in the $\Xibzstar\pim$ mass distribution (Fig.~\ref{fig:2}a),
while two new peaks are observed in the $\Xibmprime\pip$  (Fig.~\ref{fig:2}b)
and $\Xibmstar\pip$  (Fig.~\ref{fig:2}c) mass distributions,
referred to as \XibzstarstarA and \XibzstarstarB in this Letter.
For the newly observed states, the signal yields are $136 \pm 17$ for
the \XibmstarstarB resonance, $147 \pm 19$ for the \XibzstarstarA resonance and $69 \pm 14$ for the \XibzstarstarB resonance.
The three peaks are observed with local significances of
18\,$\sigma$, 15\,$\sigma$ and 9\,$\sigma$, based on the differences in log-likelihood between a fit with the signal yield fixed to zero and the default fit.
These significances are reduced to  12\,$\sigma$, 10\,$\sigma$ and 8\,$\sigma$, once systematic uncertainties on the yields are taken into account.

Different sources of systematic uncertainties are considered in the determination of the resonance parameters.
All these systematic uncertainties are summarized in Table~\ref{tab:sys}.
One of the most important contributions to the mass measurements comes from the knowledge of the momentum scale at \lhcb.
The associated systematic uncertainty is assigned as the larger of the changes in the measured parameters when the momentum scale is changed by its uncertainty, which is estimated to be $3 \times 10^{-4}$~\cite{LHCb-PAPER-2013-011}.
An additional uncertainty arises from the empirical description of the background shapes and is estimated by modeling them with alternative functions.
A third uncertainty is assigned by varying the description of the reflection components and their properties, obtained either from simulation or from data, where relevant.
Further sources of uncertainty on the measurement of the natural widths are included to describe the known differences in resolution between 
data and simulated events. Differences are expected to be within $5$\%, based on previous studies~\cite{LHCb-PAPER-2014-061,LHCb-PAPER-2017-002,LHCb-PAPER-2018-013,LHCb-PAPER-2018-032}, therefore uncertainties are estimated by varying the  resolution function width and the parametrization of the mass resolution function by that amount.
Possible uncertainties can arise from the  assumed relativistic Breit--Wigner distributions and their parameters. Lower mass states are assumed to decay with angular momentum $l=0$, while higher mass states with $l=1$.
For the newly observed states, the hypotheses assuming $l=0,2,3$ 
are tested and the largest shifts of the fitted parameters with respect to the default fit are assigned as systematic uncertainties.
A further uncertainty on the baryon mass $m_0$ is assigned due to the
limited knowledge of the \Xibm and \Xibz baryon masses~\cite{PDG2022}.
\begin{table}[t]
\caption{\small Systematic uncertainties (\mev) on the measured physical properties. The parameters
$Q_0$ and $\Gamma$ are the mean and the width of the Breit--Wigner distribution, respectively.}
\centering
\begin{tabular}{lccccccccc}
\hline
 & & \multicolumn{2}{c}{\Xibzstar}& & \multicolumn{2}{c}{\Xibmprime} & &\multicolumn{2}{c}{\Xibmstar} \\
 
Source         & & $Q_0$ &  $\Gamma$ & & $Q_0$ &  $\Gamma$ & & $Q_0$ &  $\Gamma$ \\
\hline
Momentum scale & & 0.006 &	0.001 & &  0.001 & 0.001& &	0.008&	0.001\\
Background	   & & 0.003 &	0.029 & &  0.000 & 0.006& & 0.004&	0.073\\
Reflections    & &       &        & &  0.000 & 0.000& & 0.002&  0.007\\
Resolution     & & 0.001 &	0.038 & &  0.002 & 0.027& &	0.000&	0.033\\
BW param.      & & 0.001 &	0.001 &	&  0.000 & 0.000& & 0.001&	0.002\\
Total		   & & 0.007 &  0.048 &	&  0.002 & 0.028& & 0.010&	0.081\\
\hline
\end{tabular}
\begin{tabular}{lccccccccc}
 & & \multicolumn{2}{c}{\XibmstarstarB}& & \multicolumn{2}{c}{\XibzstarstarA} & &\multicolumn{2}{c}{\XibzstarstarB} \\
Source         & & $Q_0$ &  $\Gamma$ & & $Q_0$ &  $\Gamma$ & & $Q_0$ &  $\Gamma$\\
\hline
Momentum scale & & 0.008 &	0.002 & & 0.007 &0.001 & & 0.009 & 0.006\\
Background	   & & 0.004 &	0.035 & & 0.022	&0.089 & & 0.023 & 0.025\\
Resolution	   & & 0.004 &	0.054 & & 0.001	&0.035 & & 0.006 & 0.073\\
BW param.      & & 0.016 &	0.050 & & 0.056	&0.007 & & 0.001 & 0.079\\
Total		   & & 0.019 &	0.081 & & 0.060	&0.096 & & 0.026 & 0.111\\
\hline
\end{tabular}
\label{tab:sys}
\end{table}
%

The numerical results are summarized in Table~\ref{tab:resultstot}. 
The properties of the \Xibzstar, \Xibmprime and \Xibmstar baryons are measured with world-leading precision.
For the narrow \Xibmprime state, its natural width is compatible with zero once the systematic uncertainties are considered, and an upper limit $< 0.05$\mev is estimated at 90\% confidence level.
\begin{table}[t]
\caption{\small Masses and widths of the states considered in this Letter.
The first uncertainty is statistical, the second systematic.
The third uncertainty on $m_0$ is due to limited 
knowledge of the \Xibm and \Xibz baryon masses~\cite{PDG2022}.}
\centering
\begin{tabular}{ll@{\,}  r @{$\,\pm\;$} c @{$\,\pm\;$} l  }
\hline
State & Observ. & \multicolumn{3}{c}{Value~$\mathrm{(MeV)}$}\\ 
\hline
\XibmstarstarB & $Q_0$   &    23.6 &  0.11  &  0.02  \\
               &$\Gamma$ &   0.94  &  0.30  &  0.08  \\
               &$m_0$    & 6099.74 & 0.11   &  0.02 {$\,\pm\;$}0.6 $(\Xibm)$\\
%
\hline
\XibzstarstarA &$Q_0$     &  16.20 &  0.20  &  0.06  \\ 
               & $\Gamma$ &   2.43 &  0.51  &  0.10  \\
               & $m_0$    &6087.24 &  0.20  &  0.06 {$\,\pm\;$}0.5 $(\Xibz)$\\
\hline
\XibzstarstarB & $Q_0$    &  24.32  &  0.15  &  0.03  \\
               & $\Gamma$ &   0.50  &  0.33  &  0.11  \\
               & $m_0$    & 6095.36 &  0.15  &  0.03 {$\,\pm\;$}0.5 $(\Xibz)$\\
\hline
\Xibzstar      & $Q_0$    & 15.80   &  0.02  &  0.01  \\
               & $\Gamma$ & 0.87    &  0.06  &  0.05  \\
               & $m_0$    & 5952.37 &  0.02  &  0.01 {$\,\pm\;$}0.6 $(\Xibm)$\\
\hline
\Xibmprime     & $Q_0$    &  3.66   &  0.01  &  0.00  \\
               & $\Gamma$ &  0.03   &  0.01  &  0.03  \\
               & $m_0$    & 5935.13 &  0.01  &  0.00 {$\,\pm\;$}0.5 $(\Xibz)$\\
\hline
\Xibmstar      & $Q_0$    & 24.27  &  0.03  &  0.01  \\
               & $\Gamma$ &    1.43  &  0.08  &  0.08  \\
               & $m_0$    & 5955.74  &  0.03  &  0.01 {$\,\pm\;$}0.5 $(\Xibz)$\\
\hline
\end{tabular}
\label{tab:resultstot}
\end{table}

In summary, the first observation of two new baryons \XibzstarstarA and \XibzstarstarB, with quark content $bsu$, 
is reported in the \Xibz\pip\pim final state.
Additionally, this Letter confirms the observation of the \XibmstarstarB charged state by the \cms collaboration~\cite{PhysRevLett.126.252003},
with improved significance and sensitivity on its physical parameters. This measurement uses final states with up to nine tracks, most of which are pions, showing excellent performance of the LHCb tracking, reconstruction and PID systems.
Finally, the decay mode \mbox{$\Xibz \rightarrow \Xicp\pim\pip\pim$} is observed for the first time.
The properties of the \Xibzstar, \Xibmprime and \Xibmstar baryons are measured with high precision.
Determination of the spin and parity for the new baryons is not possible given the low signal yields. However, data indicate that the \XibmstarstarB baryon decays mainly through the \Xibzstar\pim state, the \XibzstarstarA baryon mainly through the \Xibmprime\pip state, and the \XibzstarstarB baryon mainly through the \Xibmstar\pip state,
with no significant contributions to the signals from events
outside their respective $m_{\Xib\pi}$ mass windows~\cite{supplemental}.
These patterns closely resemble those observed in the \Xicz and \Xicp baryon systems~\cite{PDG2022}.
An interpretation would be that the new states are $P$-wave states
($l=1$ between the $b$ quark and the $qs$ diquark)
coupling to the $b$ quark
to give a pair of states with $J^P = {\frac{1}{2}}^-$ and
${\frac{3}{2}}^-$.
One might expect the dominant decay mode of the lighter one
to be $\Xibprime\pi$ and for the heavier one $\Xibstar\pi$.
The lighter \XibmstarstarA state could not be observed as it would likely decay primarily through the intermediate $\Xibzprime$
resonance which is below threshold to decay to $\Xibm \pi^+$. However, hints of such 
$\XibmstarstarA \rightarrow \Xibzprime(\Xibz\piz) \pim$ decay
are observed in the $\Xibz\pi^-$ spectrum as a partially reconstructed feed-down component.

\section*{Acknowledgements}
%
%
\noindent We express our gratitude to our colleagues in the CERN
accelerator departments for the excellent performance of the LHC. We
thank the technical and administrative staff at the LHCb
institutes.
We acknowledge support from CERN and from the national agencies:
CAPES, CNPq, FAPERJ and FINEP (Brazil); 
MOST and NSFC (China); 
CNRS/IN2P3 (France); 
BMBF, DFG and MPG (Germany); 
INFN (Italy); 
NWO (Netherlands); 
MNiSW and NCN (Poland); 
MCID/IFA (Romania); 
MICINN (Spain); 
SNSF and SER (Switzerland); 
NASU (Ukraine); 
STFC (United Kingdom); 
DOE NP and NSF (USA).
We acknowledge the computing resources that are provided by CERN, IN2P3
(France), KIT and DESY (Germany), INFN (Italy), SURF (Netherlands),
PIC (Spain), GridPP (United Kingdom), 
CSCS (Switzerland), IFIN-HH (Romania), CBPF (Brazil),
Polish WLCG  (Poland) and NERSC (USA).
We are indebted to the communities behind the multiple open-source
software packages on which we depend.
Individual groups or members have received support from
ARC and ARDC (Australia);
Minciencias (Colombia);
AvH Foundation (Germany);
EPLANET, Marie Sk\l{}odowska-Curie Actions, ERC and NextGenerationEU (European Union);
A*MIDEX, ANR, IPhU and Labex P2IO, and R\'{e}gion Auvergne-Rh\^{o}ne-Alpes (France);
Key Research Program of Frontier Sciences of CAS, CAS PIFI, CAS CCEPP, 
Fundamental Research Funds for the Central Universities, 
and Sci. \& Tech. Program of Guangzhou (China);
GVA, XuntaGal, GENCAT, Inditex, InTalent and Prog.~Atracci\'on Talento, CM (Spain);
SRC (Sweden);
the Leverhulme Trust, the Royal Society
 and UKRI (United Kingdom).

\addcontentsline{toc}{section}{References}
\bibliographystyle{LHCb}
\bibliography{main,standard,LHCb-PAPER,LHCb-CONF,LHCb-DP,LHCb-TDR}

\ifx\mcitethebibliography\mciteundefinedmacro
\PackageError{LHCb.bst}{mciteplus.sty has not been loaded}
{This bibstyle requires the use of the mciteplus package.}\fi
\providecommand{\href}[2]{#2}
\begin{mcitethebibliography}{10}
\mciteSetBstSublistMode{n}
\mciteSetBstMaxWidthForm{subitem}{\alph{mcitesubitemcount})}
\mciteSetBstSublistLabelBeginEnd{\mcitemaxwidthsubitemform\space}
{\relax}{\relax}

\bibitem{Gell-Mann:1964ewy}
M.~Gell-Mann, \ifthenelse{\boolean{articletitles}}{\emph{{A schematic model of
  baryons and mesons}},
  }{}\href{https://doi.org/10.1016/S0031-9163(64)92001-3}{Phys.\ Lett.\
  \textbf{8} (1964) 214}\relax
\mciteBstWouldAddEndPuncttrue
\mciteSetBstMidEndSepPunct{\mcitedefaultmidpunct}
{\mcitedefaultendpunct}{\mcitedefaultseppunct}\relax
\EndOfBibitem
\bibitem{Zweig:352337}
G.~Zweig, \ifthenelse{\boolean{articletitles}}{\emph{{An SU$_3$ model for
  strong interaction symmetry and its breaking; Version 1}}}{}
  \href{http://cds.cern.ch/record/352337}{CERN-TH-401}, CERN, Geneva,
  1964\relax
\mciteBstWouldAddEndPuncttrue
\mciteSetBstMidEndSepPunct{\mcitedefaultmidpunct}
{\mcitedefaultendpunct}{\mcitedefaultseppunct}\relax
\EndOfBibitem
\bibitem{Klempt:2009pi}
E.~Klempt and J.-M. Richard, \ifthenelse{\boolean{articletitles}}{\emph{{Baryon
  spectroscopy}}, }{}\href{https://doi.org/10.1103/RevModPhys.82.1095}{Rev.\
  Mod.\ Phys.\  \textbf{82} (2010) 1095},
  \href{http://arxiv.org/abs/0901.2055}{{\normalfont\ttfamily
  arXiv:0901.2055}}\relax
\mciteBstWouldAddEndPuncttrue
\mciteSetBstMidEndSepPunct{\mcitedefaultmidpunct}
{\mcitedefaultendpunct}{\mcitedefaultseppunct}\relax
\EndOfBibitem
\bibitem{PhysRevLett.99.052002}
CDF Collaboration, T.~Aaltonen {\em et~al.},
  \ifthenelse{\boolean{articletitles}}{\emph{{Observation and mass measurement
  of the baryon ${\ensuremath{\PXi}}_{b}^{\ensuremath{-}}$}},
  }{}\href{https://doi.org/10.1103/PhysRevLett.99.052002}{Phys.\ Rev.\ Lett.\
  \textbf{99} (2007) 052002}\relax
\mciteBstWouldAddEndPuncttrue
\mciteSetBstMidEndSepPunct{\mcitedefaultmidpunct}
{\mcitedefaultendpunct}{\mcitedefaultseppunct}\relax
\EndOfBibitem
\bibitem{CDF:2011ipk}
CDF collaboration, T.~Aaltonen {\em et~al.},
  \ifthenelse{\boolean{articletitles}}{\emph{{Observation of the $\PXi_b^0$
  baryon}}, }{}\href{https://doi.org/10.1103/PhysRevLett.107.102001}{Phys.\
  Rev.\ Lett.\  \textbf{107} (2011) 102001},
  \href{http://arxiv.org/abs/1107.4015}{{\normalfont\ttfamily
  arXiv:1107.4015}}\relax
\mciteBstWouldAddEndPuncttrue
\mciteSetBstMidEndSepPunct{\mcitedefaultmidpunct}
{\mcitedefaultendpunct}{\mcitedefaultseppunct}\relax
\EndOfBibitem
\bibitem{CMS:2012frl}
CMS collaboration, S.~Chatrchyan {\em et~al.},
  \ifthenelse{\boolean{articletitles}}{\emph{{Observation of a new $\PXi_b$
  baryon}}, }{}\href{https://doi.org/10.1103/PhysRevLett.108.252002}{Phys.\
  Rev.\ Lett.\  \textbf{108} (2012) 252002},
  \href{http://arxiv.org/abs/1204.5955}{{\normalfont\ttfamily
  arXiv:1204.5955}}\relax
\mciteBstWouldAddEndPuncttrue
\mciteSetBstMidEndSepPunct{\mcitedefaultmidpunct}
{\mcitedefaultendpunct}{\mcitedefaultseppunct}\relax
\EndOfBibitem
\bibitem{LHCb:2014nae}
LHCb collaboration, R.~Aaij {\em et~al.},
  \ifthenelse{\boolean{articletitles}}{\emph{{Observation of two new $\PXi_b^-$
  baryon resonances}},
  }{}\href{https://doi.org/10.1103/PhysRevLett.114.062004}{Phys.\ Rev.\ Lett.\
  \textbf{114} (2015) 062004},
  \href{http://arxiv.org/abs/1411.4849}{{\normalfont\ttfamily
  arXiv:1411.4849}}\relax
\mciteBstWouldAddEndPuncttrue
\mciteSetBstMidEndSepPunct{\mcitedefaultmidpunct}
{\mcitedefaultendpunct}{\mcitedefaultseppunct}\relax
\EndOfBibitem
\bibitem{LHCb-PAPER-2020-016}
LHCb collaboration, R.~Aaij {\em et~al.},
  \ifthenelse{\boolean{articletitles}}{\emph{{First branching fraction
  measurement of the suppressed decay $\Xicz \to \pim\Lc$}},
  }{}\href{https://doi.org/10.1103/PhysRevD.102.071101}{Phys.\ Rev.\
  \textbf{D102} (2020) 071101(R)},
  \href{http://arxiv.org/abs/2007.12096}{{\normalfont\ttfamily
  arXiv:2007.12096}}\relax
\mciteBstWouldAddEndPuncttrue
\mciteSetBstMidEndSepPunct{\mcitedefaultmidpunct}
{\mcitedefaultendpunct}{\mcitedefaultseppunct}\relax
\EndOfBibitem
\bibitem{PhysRevD.96.116016}
K.-L. Wang, Y.-X. Yao, X.-H. Zhong, and Q.~Zhao,
  \ifthenelse{\boolean{articletitles}}{\emph{{Strong and radiative decays of
  the low-lying $S$- and $P$-wave singly heavy baryons}},
  }{}\href{https://doi.org/10.1103/PhysRevD.96.116016}{Phys.\ Rev.\
  \textbf{D96} (2017) 116016}\relax
\mciteBstWouldAddEndPuncttrue
\mciteSetBstMidEndSepPunct{\mcitedefaultmidpunct}
{\mcitedefaultendpunct}{\mcitedefaultseppunct}\relax
\EndOfBibitem
\bibitem{PhysRevD.98.031502}
B.~Chen, K.-W. Wei, X.~Liu, and A.~Zhang,
  \ifthenelse{\boolean{articletitles}}{\emph{{Role of newly discovered
  ${{\PXi}}_{b}(6227{)}^{\ensuremath{-}}$ for constructing excited bottom
  baryon family}}, }{}\href{https://doi.org/10.1103/PhysRevD.98.031502}{Phys.\
  Rev.\  \textbf{D98} (2018) 031502}\relax
\mciteBstWouldAddEndPuncttrue
\mciteSetBstMidEndSepPunct{\mcitedefaultmidpunct}
{\mcitedefaultendpunct}{\mcitedefaultseppunct}\relax
\EndOfBibitem
\bibitem{PhysRevD.99.094016}
Y.~Kawakami and M.~Harada, \ifthenelse{\boolean{articletitles}}{\emph{{Singly
  heavy baryons with chiral partner structure in a three-flavor chiral model}},
  }{}\href{https://doi.org/10.1103/PhysRevD.99.094016}{Phys.\ Rev.\
  \textbf{D99} (2019) 094016}\relax
\mciteBstWouldAddEndPuncttrue
\mciteSetBstMidEndSepPunct{\mcitedefaultmidpunct}
{\mcitedefaultendpunct}{\mcitedefaultseppunct}\relax
\EndOfBibitem
\bibitem{PhysRevD.104.054012}
Y.~Kim, Y.-R. Liu, M.~Oka, and K.~Suzuki,
  \ifthenelse{\boolean{articletitles}}{\emph{{Heavy baryon spectrum with chiral
  multiplets of scalar and vector diquarks}},
  }{}\href{https://doi.org/10.1103/PhysRevD.104.054012}{Phys.\ Rev.\
  \textbf{D104} (2021) 054012}\relax
\mciteBstWouldAddEndPuncttrue
\mciteSetBstMidEndSepPunct{\mcitedefaultmidpunct}
{\mcitedefaultendpunct}{\mcitedefaultseppunct}\relax
\EndOfBibitem
\bibitem{th2_2021}
H.-Z. He, W.~Liang, Q.-F. Lü, and Y.-B. Dong,
  \ifthenelse{\boolean{articletitles}}{\emph{{Strong decays of the low-lying
  bottom strange baryons}},
  }{}\href{https://doi.org/10.1007/s11433-021-1704-x}{Science China Physics,
  Mechanics and Astronomy \textbf{64} (2021) }\relax
\mciteBstWouldAddEndPuncttrue
\mciteSetBstMidEndSepPunct{\mcitedefaultmidpunct}
{\mcitedefaultendpunct}{\mcitedefaultseppunct}\relax
\EndOfBibitem
\bibitem{PhysRevD.103.094003}
A.~J. Arifi, D.~Suenaga, and A.~Hosaka,
  \ifthenelse{\boolean{articletitles}}{\emph{{Relativistic corrections to
  decays of heavy baryons in the quark model}},
  }{}\href{https://doi.org/10.1103/PhysRevD.103.094003}{Phys.\ Rev.\
  \textbf{D103} (2021) 094003}\relax
\mciteBstWouldAddEndPuncttrue
\mciteSetBstMidEndSepPunct{\mcitedefaultmidpunct}
{\mcitedefaultendpunct}{\mcitedefaultseppunct}\relax
\EndOfBibitem
\bibitem{Chen:2014nyo}
B.~Chen, K.-W. Wei, and A.~Zhang,
  \ifthenelse{\boolean{articletitles}}{\emph{{Assignments of $\PLambda_Q$ and
  $\PXi_Q$ baryons in the heavy quark-light diquark picture}},
  }{}\href{https://doi.org/10.1140/epja/i2015-15082-3}{Eur.\ Phys.\ J.\
  \textbf{A51} (2015) 82},
  \href{http://arxiv.org/abs/1406.6561}{{\normalfont\ttfamily
  arXiv:1406.6561}}\relax
\mciteBstWouldAddEndPuncttrue
\mciteSetBstMidEndSepPunct{\mcitedefaultmidpunct}
{\mcitedefaultendpunct}{\mcitedefaultseppunct}\relax
\EndOfBibitem
\bibitem{PhysRevD.84.014025}
D.~Ebert, R.~N. Faustov, and V.~O. Galkin,
  \ifthenelse{\boolean{articletitles}}{\emph{{Spectroscopy and Regge
  trajectories of heavy baryons in the relativistic quark-diquark picture}},
  }{}\href{https://doi.org/10.1103/PhysRevD.84.014025}{Phys.\ Rev.\
  \textbf{D84} (2011) 014025}\relax
\mciteBstWouldAddEndPuncttrue
\mciteSetBstMidEndSepPunct{\mcitedefaultmidpunct}
{\mcitedefaultendpunct}{\mcitedefaultseppunct}\relax
\EndOfBibitem
\bibitem{Roberts:2007ni}
W.~Roberts and M.~Pervin, \ifthenelse{\boolean{articletitles}}{\emph{{Heavy
  baryons in a quark model}},
  }{}\href{https://doi.org/10.1142/S0217751X08041219}{Int.\ J.\ Mod.\ Phys.\
  \textbf{A23} (2008) 2817},
  \href{http://arxiv.org/abs/0711.2492}{{\normalfont\ttfamily
  arXiv:0711.2492}}\relax
\mciteBstWouldAddEndPuncttrue
\mciteSetBstMidEndSepPunct{\mcitedefaultmidpunct}
{\mcitedefaultendpunct}{\mcitedefaultseppunct}\relax
\EndOfBibitem
\bibitem{PhysRevLett.126.252003}
CMS collaboration, A.~M. Sirunyan {\em et~al.},
  \ifthenelse{\boolean{articletitles}}{\emph{{Observation of a new excited
  beauty strange Baryon Decaying to ${\PXi}_{b}^{-}{\pi}^{+}{\pi}^{-}$}},
  }{}\href{https://doi.org/10.1103/PhysRevLett.126.252003}{Phys.\ Rev.\ Lett.\
  \textbf{126} (2021) 252003}\relax
\mciteBstWouldAddEndPuncttrue
\mciteSetBstMidEndSepPunct{\mcitedefaultmidpunct}
{\mcitedefaultendpunct}{\mcitedefaultseppunct}\relax
\EndOfBibitem
\bibitem{LHCb-DP-2008-001}
LHCb collaboration, A.~A. Alves~Jr.\ {\em et~al.},
  \ifthenelse{\boolean{articletitles}}{\emph{{The \lhcb detector at the LHC}},
  }{}\href{https://doi.org/10.1088/1748-0221/3/08/S08005}{JINST \textbf{3}
  (2008) S08005}\relax
\mciteBstWouldAddEndPuncttrue
\mciteSetBstMidEndSepPunct{\mcitedefaultmidpunct}
{\mcitedefaultendpunct}{\mcitedefaultseppunct}\relax
\EndOfBibitem
\bibitem{LHCb-DP-2014-002}
LHCb collaboration, R.~Aaij {\em et~al.},
  \ifthenelse{\boolean{articletitles}}{\emph{{LHCb detector performance}},
  }{}\href{https://doi.org/10.1142/S0217751X15300227}{Int.\ J.\ Mod.\ Phys.\
  \textbf{A30} (2015) 1530022},
  \href{http://arxiv.org/abs/1412.6352}{{\normalfont\ttfamily
  arXiv:1412.6352}}\relax
\mciteBstWouldAddEndPuncttrue
\mciteSetBstMidEndSepPunct{\mcitedefaultmidpunct}
{\mcitedefaultendpunct}{\mcitedefaultseppunct}\relax
\EndOfBibitem
\bibitem{LHCb-DP-2014-001}
R.~Aaij {\em et~al.}, \ifthenelse{\boolean{articletitles}}{\emph{{Performance
  of the LHCb Vertex Locator}},
  }{}\href{https://doi.org/10.1088/1748-0221/9/09/P09007}{JINST \textbf{9}
  (2014) P09007}, \href{http://arxiv.org/abs/1405.7808}{{\normalfont\ttfamily
  arXiv:1405.7808}}\relax
\mciteBstWouldAddEndPuncttrue
\mciteSetBstMidEndSepPunct{\mcitedefaultmidpunct}
{\mcitedefaultendpunct}{\mcitedefaultseppunct}\relax
\EndOfBibitem
\bibitem{LHCb-DP-2013-003}
R.~Arink {\em et~al.}, \ifthenelse{\boolean{articletitles}}{\emph{{Performance
  of the LHCb Outer Tracker}},
  }{}\href{https://doi.org/10.1088/1748-0221/9/01/P01002}{JINST \textbf{9}
  (2014) P01002}, \href{http://arxiv.org/abs/1311.3893}{{\normalfont\ttfamily
  arXiv:1311.3893}}\relax
\mciteBstWouldAddEndPuncttrue
\mciteSetBstMidEndSepPunct{\mcitedefaultmidpunct}
{\mcitedefaultendpunct}{\mcitedefaultseppunct}\relax
\EndOfBibitem
\bibitem{LHCb-DP-2017-001}
P.~d'Argent {\em et~al.}, \ifthenelse{\boolean{articletitles}}{\emph{{Improved
  performance of the LHCb Outer Tracker in LHC Run 2}},
  }{}\href{https://doi.org/10.1088/1748-0221/12/11/P11016}{JINST \textbf{12}
  (2017) P11016}, \href{http://arxiv.org/abs/1708.00819}{{\normalfont\ttfamily
  arXiv:1708.00819}}\relax
\mciteBstWouldAddEndPuncttrue
\mciteSetBstMidEndSepPunct{\mcitedefaultmidpunct}
{\mcitedefaultendpunct}{\mcitedefaultseppunct}\relax
\EndOfBibitem
\bibitem{LHCb-DP-2012-003}
M.~Adinolfi {\em et~al.},
  \ifthenelse{\boolean{articletitles}}{\emph{{Performance of the \lhcb RICH
  detector at the LHC}},
  }{}\href{https://doi.org/10.1140/epjc/s10052-013-2431-9}{Eur.\ Phys.\ J.\
  \textbf{C73} (2013) 2431},
  \href{http://arxiv.org/abs/1211.6759}{{\normalfont\ttfamily
  arXiv:1211.6759}}\relax
\mciteBstWouldAddEndPuncttrue
\mciteSetBstMidEndSepPunct{\mcitedefaultmidpunct}
{\mcitedefaultendpunct}{\mcitedefaultseppunct}\relax
\EndOfBibitem
\bibitem{Sjostrand:2007gs}
T.~Sj\"{o}strand, S.~Mrenna, and P.~Skands,
  \ifthenelse{\boolean{articletitles}}{\emph{{A brief introduction to PYTHIA
  8.1}}, }{}\href{https://doi.org/10.1016/j.cpc.2008.01.036}{Comput.\ Phys.\
  Commun.\  \textbf{178} (2008) 852},
  \href{http://arxiv.org/abs/0710.3820}{{\normalfont\ttfamily
  arXiv:0710.3820}}\relax
\mciteBstWouldAddEndPuncttrue
\mciteSetBstMidEndSepPunct{\mcitedefaultmidpunct}
{\mcitedefaultendpunct}{\mcitedefaultseppunct}\relax
\EndOfBibitem
\bibitem{LHCb:2011dpk}
I.~Belyaev {\em et~al.}, \ifthenelse{\boolean{articletitles}}{\emph{{Handling
  of the generation of primary events in Gauss, the LHCb simulation
  framework}}, }{}\href{https://doi.org/10.1088/1742-6596/331/3/032047}{J.\
  Phys.\ Conf.\ Ser.\  \textbf{331} (2011) 032047}\relax
\mciteBstWouldAddEndPuncttrue
\mciteSetBstMidEndSepPunct{\mcitedefaultmidpunct}
{\mcitedefaultendpunct}{\mcitedefaultseppunct}\relax
\EndOfBibitem
\bibitem{Lange:2001uf}
D.~J. Lange, \ifthenelse{\boolean{articletitles}}{\emph{{The EvtGen particle
  decay simulation package}},
  }{}\href{https://doi.org/10.1016/S0168-9002(01)00089-4}{Nucl.\ Instrum.\
  Meth.\  \textbf{A462} (2001) 152}\relax
\mciteBstWouldAddEndPuncttrue
\mciteSetBstMidEndSepPunct{\mcitedefaultmidpunct}
{\mcitedefaultendpunct}{\mcitedefaultseppunct}\relax
\EndOfBibitem
\bibitem{Allison:2006ve}
Geant4 collaboration, J.~Allison {\em et~al.},
  \ifthenelse{\boolean{articletitles}}{\emph{{Geant4 developments and
  applications}}, }{}\href{https://doi.org/10.1109/TNS.2006.869826}{IEEE
  Trans.\ Nucl.\ Sci.\  \textbf{53} (2006) 270}\relax
\mciteBstWouldAddEndPuncttrue
\mciteSetBstMidEndSepPunct{\mcitedefaultmidpunct}
{\mcitedefaultendpunct}{\mcitedefaultseppunct}\relax
\EndOfBibitem
\bibitem{Clemencic_2011}
M.~Clemencic {\em et~al.}, \ifthenelse{\boolean{articletitles}}{\emph{{The LHCb
  Simulation Application, Gauss: Design, Evolution and Experience}},
  }{}\href{https://doi.org/10.1088/1742-6596/331/3/032023}{J.\ Phys.\ Conf.\
  Ser.\  \textbf{331} (2011) 032023}\relax
\mciteBstWouldAddEndPuncttrue
\mciteSetBstMidEndSepPunct{\mcitedefaultmidpunct}
{\mcitedefaultendpunct}{\mcitedefaultseppunct}\relax
\EndOfBibitem
\bibitem{PDG2022}
Particle Data Group, R.~L. Workman {\em et~al.},
  \ifthenelse{\boolean{articletitles}}{\emph{{\href{http://pdg.lbl.gov/}{Review
  of particle physics}}}, }{}\href{https://doi.org/10.1093/ptep/ptac097}{Prog.\
  Theor.\ Exp.\ Phys.\  \textbf{2022} (2022) 083C01}\relax
\mciteBstWouldAddEndPuncttrue
\mciteSetBstMidEndSepPunct{\mcitedefaultmidpunct}
{\mcitedefaultendpunct}{\mcitedefaultseppunct}\relax
\EndOfBibitem
\bibitem{DeCian:2255039}
M.~De~Cian, S.~Farry, P.~Seyfert, and S.~Stahl,
  \ifthenelse{\boolean{articletitles}}{\emph{{Fast neural-net based fake track
  rejection in the LHCb reconstruction}}, }{}
  \href{http://cdsweb.cern.ch/search?p=LHCb-PUB-2017-011&f=reportnumber&action_search=Search&c=LHCb+Notes}
  {LHCb-PUB-2017-011}, 2017\relax
\mciteBstWouldAddEndPuncttrue
\mciteSetBstMidEndSepPunct{\mcitedefaultmidpunct}
{\mcitedefaultendpunct}{\mcitedefaultseppunct}\relax
\EndOfBibitem
\bibitem{LHCb-DP-2018-001}
R.~Aaij {\em et~al.}, \ifthenelse{\boolean{articletitles}}{\emph{{Selection and
  processing of calibration samples to measure the particle identification
  performance of the LHCb experiment in Run 2}},
  }{}\href{https://doi.org/10.1140/epjti/s40485-019-0050-z}{Eur.\ Phys.\ J.\
  Tech.\ Instr.\  \textbf{6} (2019) 1},
  \href{http://arxiv.org/abs/1803.00824}{{\normalfont\ttfamily
  arXiv:1803.00824}}\relax
\mciteBstWouldAddEndPuncttrue
\mciteSetBstMidEndSepPunct{\mcitedefaultmidpunct}
{\mcitedefaultendpunct}{\mcitedefaultseppunct}\relax
\EndOfBibitem
\bibitem{ROE2005577}
B.~P. Roe {\em et~al.}, \ifthenelse{\boolean{articletitles}}{\emph{{Boosted
  decision trees as an alternative to artificial neural networks for particle
  identification}},
  }{}\href{https://doi.org/https://doi.org/10.1016/j.nima.2004.12.018}{Nucl.\
  Instrum.\ Meth.\  \textbf{543} (2005) 577}\relax
\mciteBstWouldAddEndPuncttrue
\mciteSetBstMidEndSepPunct{\mcitedefaultmidpunct}
{\mcitedefaultendpunct}{\mcitedefaultseppunct}\relax
\EndOfBibitem
\bibitem{supplemental}
\ifthenelse{\boolean{articletitles}}{\emph{{See Supplemental Material at [URL
  will be inserted by publisher] for further details}}, }{}\relax
\mciteBstWouldAddEndPuncttrue
\mciteSetBstMidEndSepPunct{\mcitedefaultmidpunct}
{\mcitedefaultendpunct}{\mcitedefaultseppunct}\relax
\EndOfBibitem
\bibitem{LHCb-PAPER-2019-042}
LHCb collaboration, R.~Aaij {\em et~al.},
  \ifthenelse{\boolean{articletitles}}{\emph{{First observation of excited
  $\Omegares_b^-$ states}},
  }{}\href{https://doi.org/10.1103/PhysRevLett.124.082002}{Phys.\ Rev.\ Lett.\
  \textbf{124} (2020) 082002},
  \href{http://arxiv.org/abs/2001.00851}{{\normalfont\ttfamily
  arXiv:2001.00851}}\relax
\mciteBstWouldAddEndPuncttrue
\mciteSetBstMidEndSepPunct{\mcitedefaultmidpunct}
{\mcitedefaultendpunct}{\mcitedefaultseppunct}\relax
\EndOfBibitem
\bibitem{Hulsbergen:2005pu}
W.~D. Hulsbergen, \ifthenelse{\boolean{articletitles}}{\emph{{Decay chain
  fitting with a Kalman filter}},
  }{}\href{https://doi.org/10.1016/j.nima.2005.06.078}{Nucl.\ Instrum.\ Meth.\
  \textbf{A552} (2005) 566},
  \href{http://arxiv.org/abs/physics/0503191}{{\normalfont\ttfamily
  arXiv:physics/0503191}}\relax
\mciteBstWouldAddEndPuncttrue
\mciteSetBstMidEndSepPunct{\mcitedefaultmidpunct}
{\mcitedefaultendpunct}{\mcitedefaultseppunct}\relax
\EndOfBibitem
\bibitem{Jackson1964}
J.~D. Jackson, \ifthenelse{\boolean{articletitles}}{\emph{{Remarks on the
  phenomenological analysis of resonances}},
  }{}\href{https://doi.org/https://doi.org/10.1007/BF02750563}{Nuovo Cim.\
  \textbf{34} (1964) 1644–1666}\relax
\mciteBstWouldAddEndPuncttrue
\mciteSetBstMidEndSepPunct{\mcitedefaultmidpunct}
{\mcitedefaultendpunct}{\mcitedefaultseppunct}\relax
\EndOfBibitem
\bibitem{Blatt:1952ije}
J.~M. Blatt and V.~F. Weisskopf, {\em {Theoretical nuclear physics}},
  \href{https://doi.org/10.1007/978-1-4612-9959-2}{ Springer, New York,
  1952}\relax
\mciteBstWouldAddEndPuncttrue
\mciteSetBstMidEndSepPunct{\mcitedefaultmidpunct}
{\mcitedefaultendpunct}{\mcitedefaultseppunct}\relax
\EndOfBibitem
\bibitem{Skwarnicki:1986xj}
T.~Skwarnicki, {\em {A study of the radiative cascade transitions between the
  Upsilon-prime and Upsilon resonances}}, PhD thesis, Institute of Nuclear
  Physics, Krakow, 1986,
  {\href{http://inspirehep.net/record/230779/}{DESY-F31-86-02}}\relax
\mciteBstWouldAddEndPuncttrue
\mciteSetBstMidEndSepPunct{\mcitedefaultmidpunct}
{\mcitedefaultendpunct}{\mcitedefaultseppunct}\relax
\EndOfBibitem
\bibitem{LHCb-PAPER-2013-011}
LHCb collaboration, R.~Aaij {\em et~al.},
  \ifthenelse{\boolean{articletitles}}{\emph{{Precision measurement of \D meson
  mass differences}}, }{}\href{https://doi.org/10.1007/JHEP06(2013)065}{JHEP
  \textbf{06} (2013) 065},
  \href{http://arxiv.org/abs/1304.6865}{{\normalfont\ttfamily
  arXiv:1304.6865}}\relax
\mciteBstWouldAddEndPuncttrue
\mciteSetBstMidEndSepPunct{\mcitedefaultmidpunct}
{\mcitedefaultendpunct}{\mcitedefaultseppunct}\relax
\EndOfBibitem
\bibitem{LHCb-PAPER-2014-061}
LHCb collaboration, R.~Aaij {\em et~al.},
  \ifthenelse{\boolean{articletitles}}{\emph{{Observation of two new \Xibm
  baryon resonances}},
  }{}\href{https://doi.org/10.1103/PhysRevLett.114.062004}{Phys.\ Rev.\ Lett.\
  \textbf{114} (2015) 062004},
  \href{http://arxiv.org/abs/1411.4849}{{\normalfont\ttfamily
  arXiv:1411.4849}}\relax
\mciteBstWouldAddEndPuncttrue
\mciteSetBstMidEndSepPunct{\mcitedefaultmidpunct}
{\mcitedefaultendpunct}{\mcitedefaultseppunct}\relax
\EndOfBibitem
\bibitem{LHCb-PAPER-2017-002}
LHCb collaboration, R.~Aaij {\em et~al.},
  \ifthenelse{\boolean{articletitles}}{\emph{{Observation of five new narrow
  $\Omegac$ states decaying to $\Xicp\Km$}},
  }{}\href{https://doi.org/10.1103/PhysRevLett.118.182001}{Phys.\ Rev.\ Lett.\
  \textbf{118} (2017) 182001},
  \href{http://arxiv.org/abs/1703.04639}{{\normalfont\ttfamily
  arXiv:1703.04639}}\relax
\mciteBstWouldAddEndPuncttrue
\mciteSetBstMidEndSepPunct{\mcitedefaultmidpunct}
{\mcitedefaultendpunct}{\mcitedefaultseppunct}\relax
\EndOfBibitem
\bibitem{LHCb-PAPER-2018-013}
LHCb collaboration, R.~Aaij {\em et~al.},
  \ifthenelse{\boolean{articletitles}}{\emph{{Observation of a new \Xibm
  resonance}}, }{}\href{https://doi.org/10.1103/PhysRevLett.121.072002}{Phys.\
  Rev.\ Lett.\  \textbf{121} (2018) 072002},
  \href{http://arxiv.org/abs/1805.09418}{{\normalfont\ttfamily
  arXiv:1805.09418}}\relax
\mciteBstWouldAddEndPuncttrue
\mciteSetBstMidEndSepPunct{\mcitedefaultmidpunct}
{\mcitedefaultendpunct}{\mcitedefaultseppunct}\relax
\EndOfBibitem
\bibitem{LHCb-PAPER-2018-032}
LHCb collaboration, R.~Aaij {\em et~al.},
  \ifthenelse{\boolean{articletitles}}{\emph{{Observation of two resonances in
  the $\Lb\pipm$ systems and precise measurement of $\Sigmares_b^\pm$ and
  $\Sigmares_b^{\ast\pm}$ properties}},
  }{}\href{https://doi.org/10.1103/PhysRevLett.122.012001}{Phys.\ Rev.\ Lett.\
  \textbf{122} (2019) 012001},
  \href{http://arxiv.org/abs/1809.07752}{{\normalfont\ttfamily
  arXiv:1809.07752}}\relax
\mciteBstWouldAddEndPuncttrue
\mciteSetBstMidEndSepPunct{\mcitedefaultmidpunct}
{\mcitedefaultendpunct}{\mcitedefaultseppunct}\relax
\EndOfBibitem
\end{mcitethebibliography}
\cleardoublepage
\section*{Supplemental material}
\begin{figure}[tbhp]
\centering
    \ifthenelse{\boolean{prl}}
    {\includegraphics[width=1.00\linewidth]{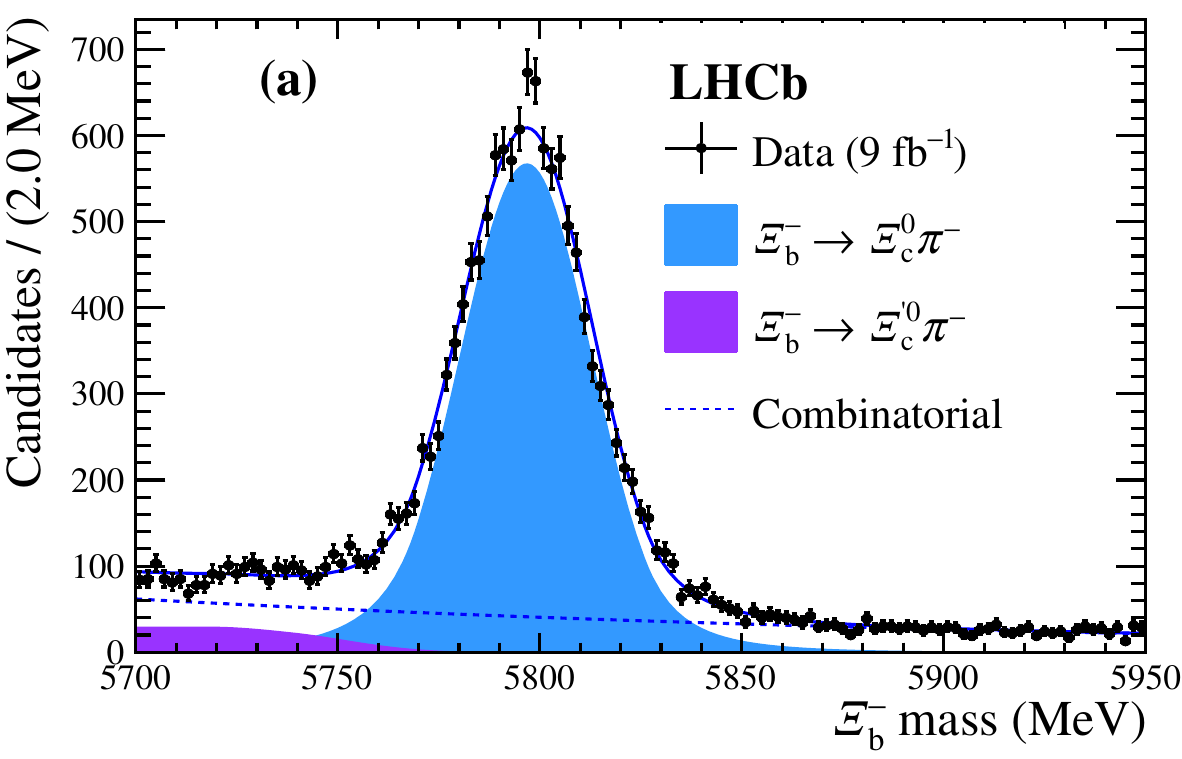}}
    {\includegraphics[width=0.46\linewidth]{fig3a.pdf}}
    \ifthenelse{\boolean{prl}}
    {\includegraphics[width=1.00\linewidth]{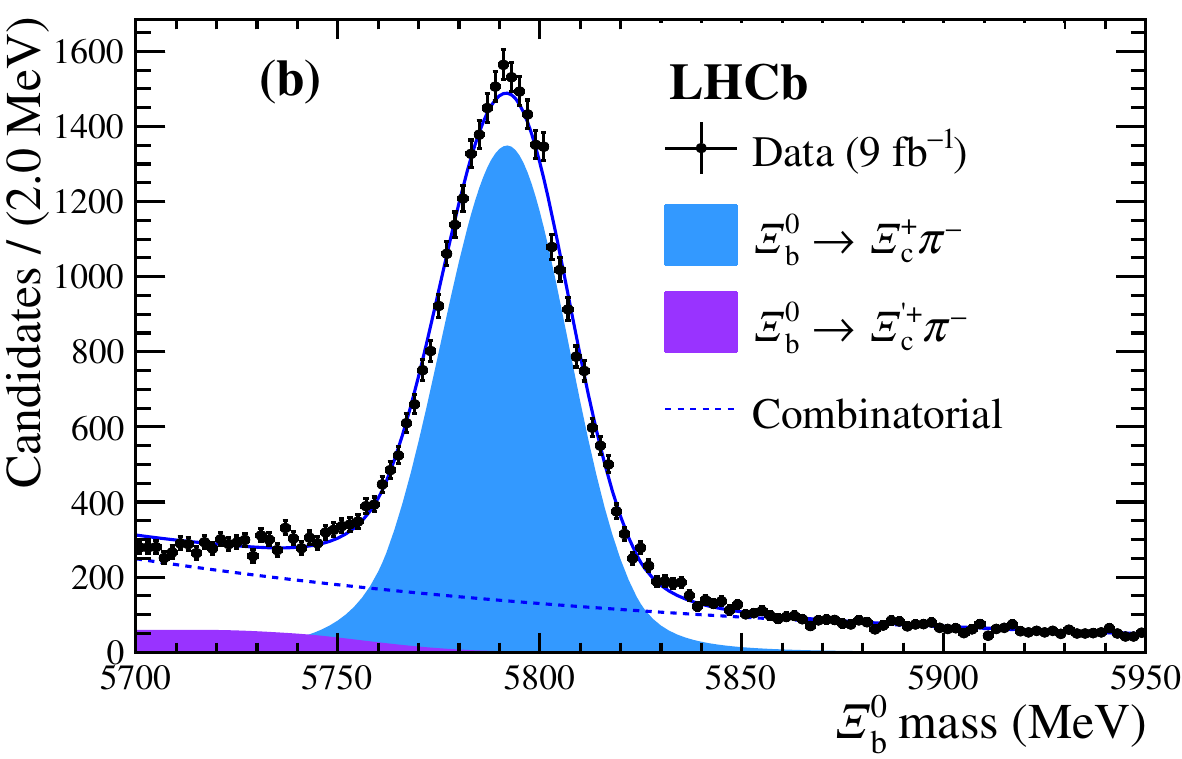}}
    {\includegraphics[width=0.46\linewidth]{fig3b.pdf}}
    \ifthenelse{\boolean{prl}}
    {\includegraphics[width=1.00\linewidth]{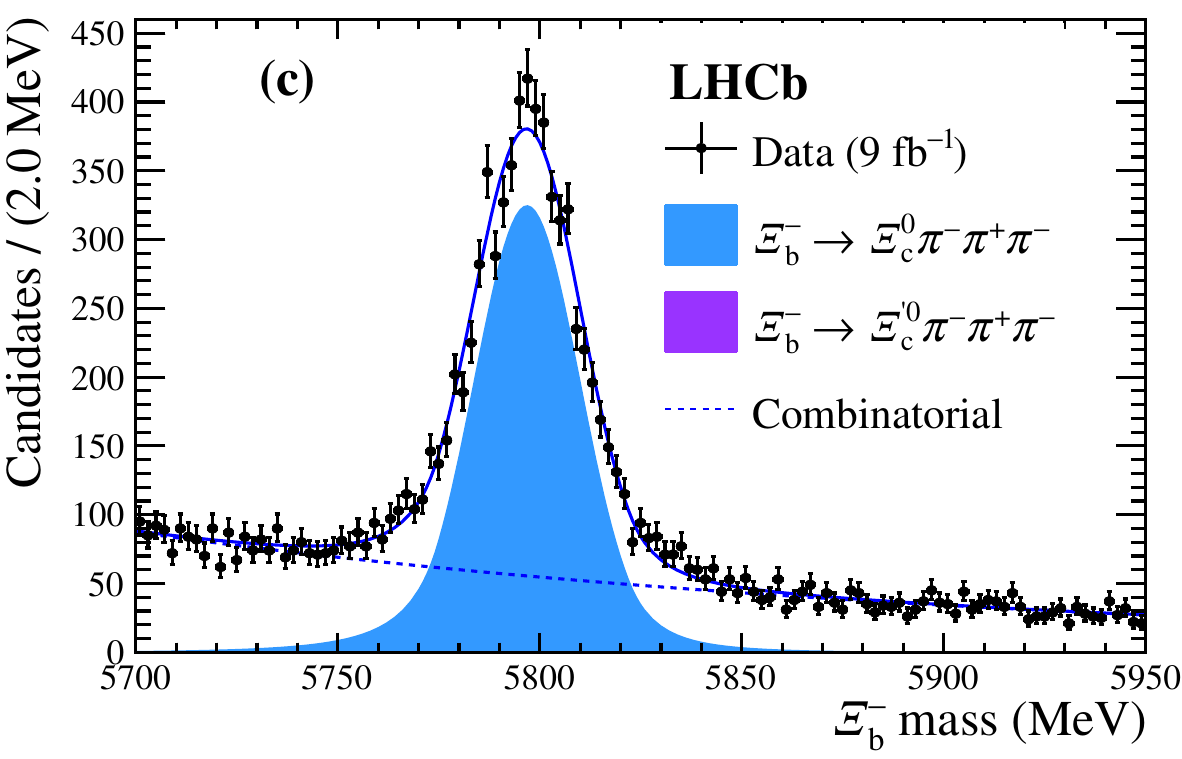}}
    {\includegraphics[width=0.46\linewidth]{fig3c.pdf}}
    \ifthenelse{\boolean{prl}}
    {\includegraphics[width=1.00\linewidth]{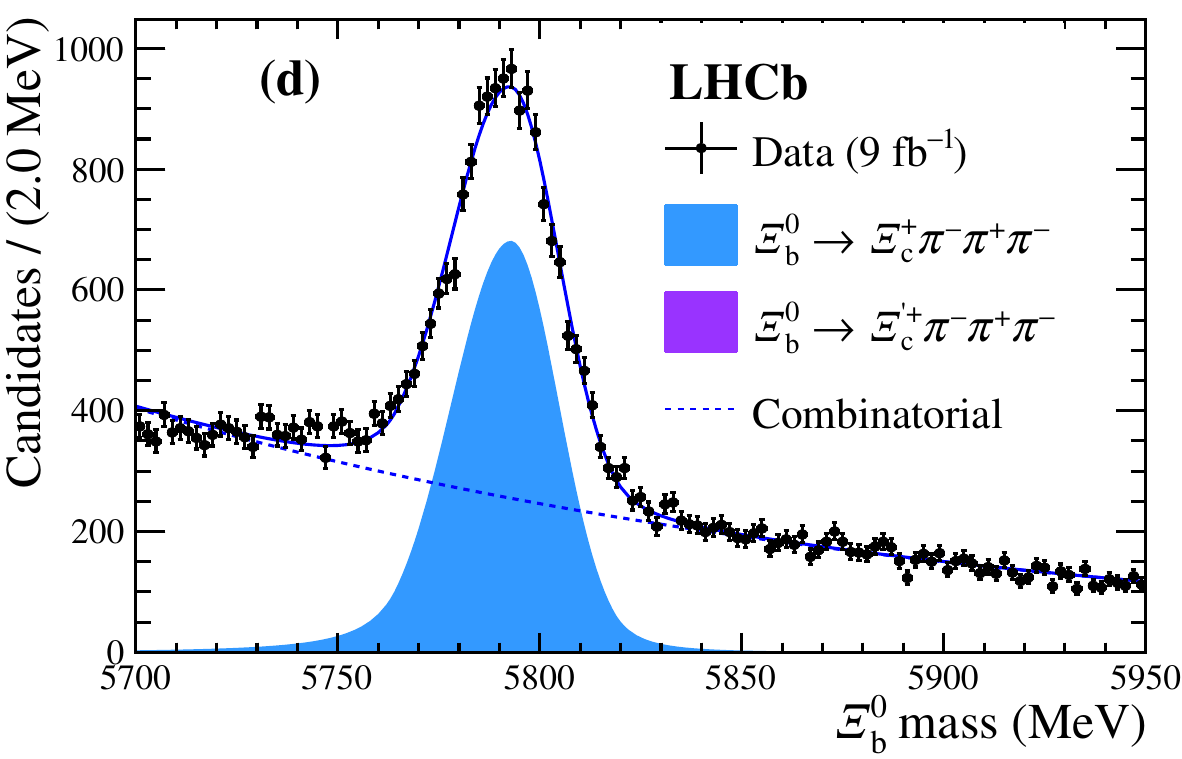}}
    {\includegraphics[width=0.46\linewidth]{fig3d.pdf}}
    \caption{\small Mass distributions of the
    (a) $\Xibm\rightarrow \Xicz\pim$,
    (b) $\Xibz\rightarrow \Xicp\pim$,
    (c) $\Xibm\rightarrow \Xicz\pim\pip\pim$ and
    (d) $\Xibz\rightarrow \Xicp\pim\pip\pim$ 
    selected signal candidates.
    \label{fig:3}}
\end{figure}

Figure~\ref{fig:3} shows the mass distributions of the selected 
\Xibm and \Xibz signal candidates.
These distributions are fitted with an asymmetric Gaussian-like function
with power-law tails for the signal component, an exponential function for the combinatorial background and an empirical function for partially reconstructed backgrounds.
The fitted signal yields are 
$12\,020 \pm  140$ for $\Xibm\rightarrow \Xicz\pim$,
$27\,300 \pm  210$ for $\Xibz\rightarrow \Xicp\pim$,
$ 5\,680 \pm  100$ for $\Xibm\rightarrow \Xicz\pim\pip\pim$,
$11\,690 \pm  180$ for $\Xibz\rightarrow \Xicp\pim\pip\pim$ decays.

\begin{figure}[tbhp]
\centering
\ifthenelse{\boolean{prl}}
{\includegraphics[width=1.00\linewidth]{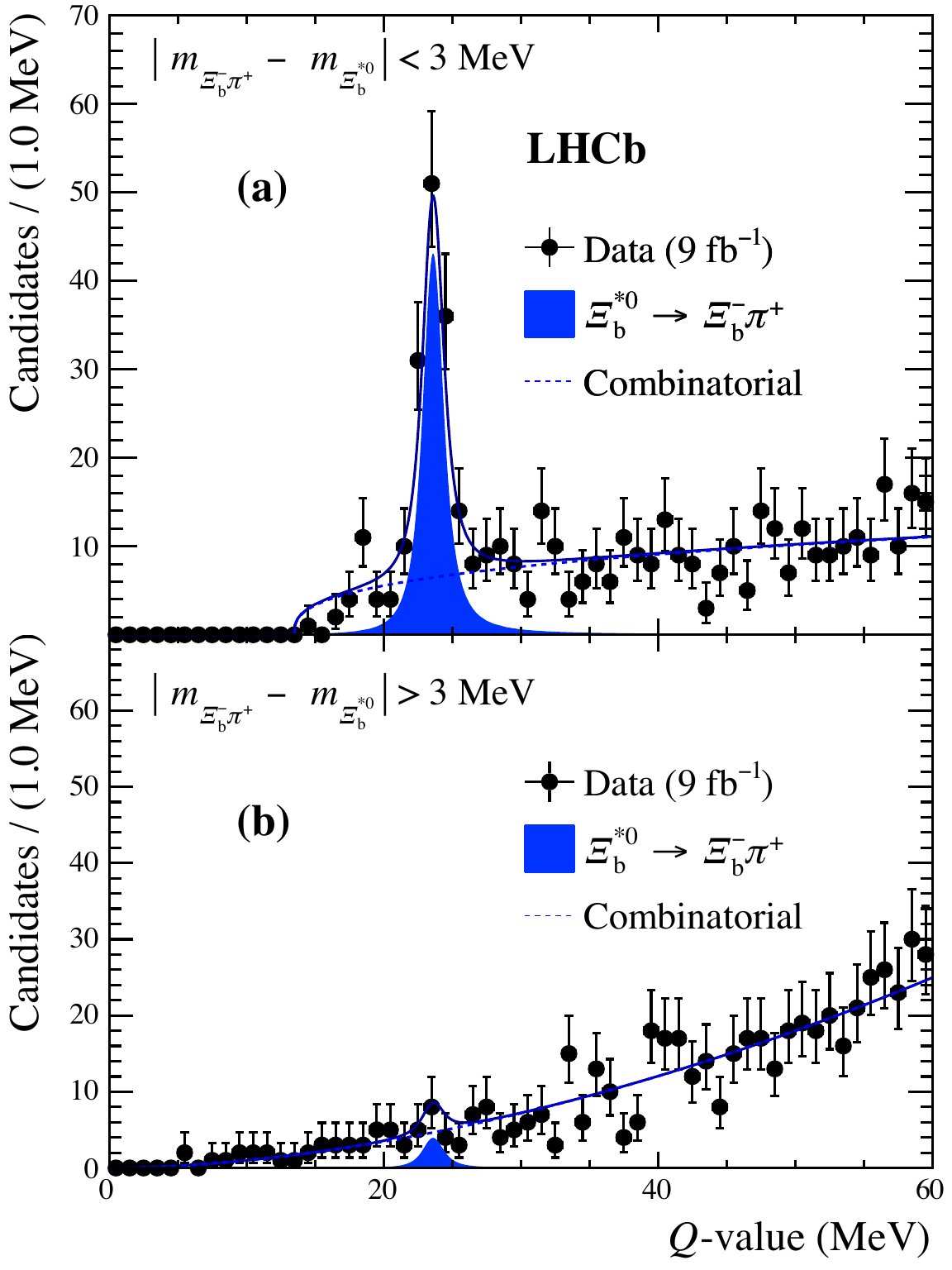}}
{\includegraphics[width=0.60\linewidth]{fig4.pdf}}
\caption{\small
Distributions of the mass difference $Q = m_{\Xib\pi\pi} - m_{\Xib} - 2 m_{\pi}$ for selected \Xibm\pip\pim candidates, (a) in the \Xibzstar mass window and (b) outside the \Xibzstar mass window. The results of the simultaneous fit are superimposed.\label{fig:4}}
\end{figure}

\begin{figure}[!ht]
\centering
\ifthenelse{\boolean{prl}}
{\includegraphics[width=1.00\linewidth]{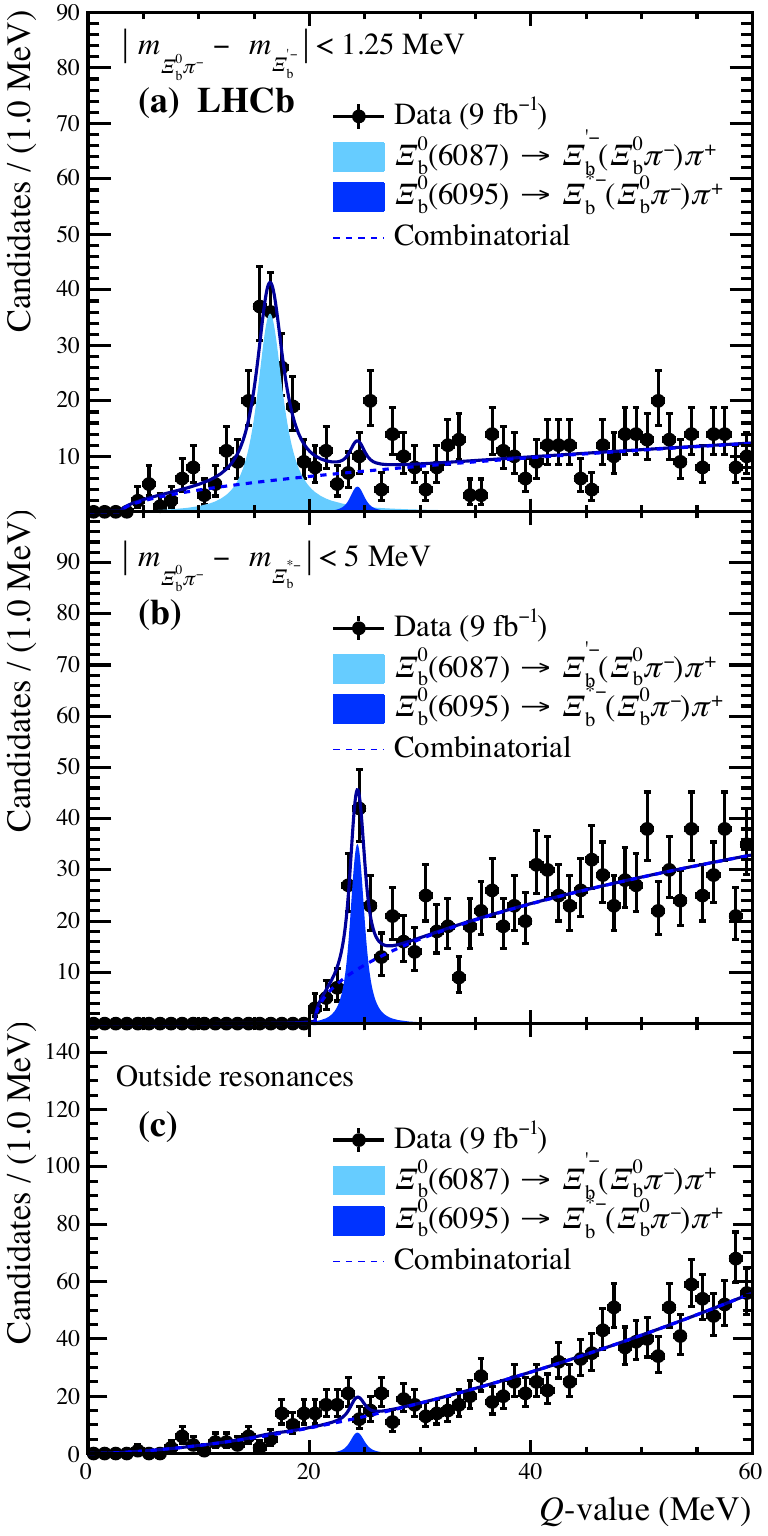}}
{\includegraphics[width=0.60\linewidth]{fig5.pdf}}
\caption{Distributions of the mass difference $Q = m_{\Xib\pi\pi} - m_{\Xib} - 2 m_{\pi}$ for selected \Xibz\pip\pim candidates, (a) in the \Xibmprime mass window, (b) in the \Xibmstar mass window and (c) outside the mass windows of the intermediate resonances. The results of the simultaneous fit are superimposed.\label{fig:5}}
\end{figure}

Figures~\ref{fig:4} and~\ref{fig:5} show the mass-difference ($Q = m_{\Xib\pi\pi} - m_{\Xib} - 2 m_{\pi}$) distributions inside and outside the mass windows corresponding to the known $\Xib\pi$ intermediate resonances. Simultaneous fits are performed using the fit models presented in the Letter, with the resonance parameters fixed to their nominal values. Yields are allowed to vary in the fit. The observed suppression of decays not proceeding through the respective intermediate resonances confirms the decay pattern described in the text.

\clearpage

\newpage
\centerline
{\large\bf LHCb collaboration}
\begin
{flushleft}
\small
R.~Aaij$^{32}$\lhcborcid{0000-0003-0533-1952},
A.S.W.~Abdelmotteleb$^{51}$\lhcborcid{0000-0001-7905-0542},
C.~Abellan~Beteta$^{45}$,
F.~Abudin{\'e}n$^{51}$\lhcborcid{0000-0002-6737-3528},
T.~Ackernley$^{55}$\lhcborcid{0000-0002-5951-3498},
B.~Adeva$^{41}$\lhcborcid{0000-0001-9756-3712},
M.~Adinolfi$^{49}$\lhcborcid{0000-0002-1326-1264},
P.~Adlarson$^{77}$\lhcborcid{0000-0001-6280-3851},
H.~Afsharnia$^{9}$,
C.~Agapopoulou$^{43}$\lhcborcid{0000-0002-2368-0147}
% ...

{\footnotesize \it

% ...
$^{9}$Universit{\'e} Clermont Auvergne, CNRS/IN2P3, LPC, Clermont-Ferrand, France\\
% ...
$^{32}$Nikhef National Institute for Subatomic Physics, Amsterdam, Netherlands\\
% ...
$^{41}$Instituto Galego de F{\'\i}sica de Altas Enerx{\'\i}as (IGFAE), Universidade de Santiago de Compostela, Santiago de Compostela, Spain\\
% ...
$^{43}$European Organization for Nuclear Research (CERN), Geneva, Switzerland\\
% ...
$^{45}$Physik-Institut, Universit{\"a}t Z{\"u}rich, Z{\"u}rich, Switzerland\\
% ...
$^{49}$H.H. Wills Physics Laboratory, University of Bristol, Bristol, United Kingdom\\
% ...
$^{51}$Department of Physics, University of Warwick, Coventry, United Kingdom\\
% ...
$^{54}$School of Physics and Astronomy, University of Glasgow, Glasgow, United Kingdom\\
$^{55}$Oliver Lodge Laboratory, University of Liverpool, Liverpool, United Kingdom\\
% ...
$^{77}$Department of Physics and Astronomy, Uppsala University, Uppsala, Sweden, associated to $^{54}$\\
% ...
}
\end{flushleft}

\end{document}